\documentclass[sigconf,9pt,nonacm]{acmart}
\usepackage{subcaption}
\usepackage{soul}
\usepackage{siunitx}
\usepackage{multirow} 
\usepackage{ulem}

\AtBeginDocument{%
  }

\newcommand\para[1]{{\vspace{3pt} \bf \noindent #1}}

\newcounter{reviewcounter}
\newcommand{\new}[1]{{\color{black}#1}}

\usepackage{balance}

\RequirePackage{ifthen}
\newboolean{editmode}
\setboolean{editmode}{true}

\RequirePackage{xcolor}
\definecolor{addcolor}{rgb}{0,0.6,0} 
\definecolor{removecolor}{rgb}{1,0,0} 
\definecolor{updatecolor}{rgb}{0,0,1} 

\ifthenelse{\boolean{editmode}}{%
  
  \newcommand{\remove}[1]{\textcolor{removecolor}{\textbf{[Removed]} #1}}
  
}{%
  
  \newcommand{\remove}[1]{}
  
}

\begin{document}
\settopmatter{printfolios=true}



\title{\systemname{}: Ego-Centric Fine-Grained Daily Log with Ubiquitous Wearables}

\author{Lixing He}
\email{1155170464@link.cuhk.edu.hk}
\affiliation{%
  \institution{The Chinese University of Hong Kong}
  \city{Hong Kong}
  \country{China}
}

\author{Bufang Yang}
\email{bfyang@link.cuhk.edu.hk}
\affiliation{%
  \institution{The Chinese University of Hong Kong}
  \city{Hong Kong}
  \country{China}
}

\author{Di Duan}
\email{diduan@ie.cuhk.edu.hk}
\affiliation{%
  \institution{The Chinese University of Hong Kong}
  \city{Hong Kong}
  \country{China}
}

\author{Zhenyu Yan}
\email{zyyan@ie.cuhk.edu.hk}
\affiliation{%
  \institution{The Chinese University of Hong Kong}
  \city{Hong Kong}
  \country{China}
}

\newcommand{\systemname}{EgoLog}
\newcommand{\fig}{Fig.~}

\begin{abstract}
Despite advances in human activity recognition (HAR) with different modalities, a precise, robust, and accurate daily log system is not yet available. Current solutions primarily rely on controlled, lab-based data collection, which limits their real-world applicability.
The challenges towards a fine-grained daily log are 1) contextual awareness, 2) spatial awareness, and 3) effective fusion of multi-modal sensor data.
To solve them, we propose \systemname{}, which integrates effective audio-IMU fusion for daily log with ubiquitous wearables.
Our approach first fuses audio and IMU data from two perspectives: temporal understanding and spatial understanding. We extract scenario-level features and aggregate them in the time dimension, while using motion compensation to enhance the performance of sound source localization.
The knowledge obtained from these steps is then integrated into a multi-modal HAR framework. Here, the scenario provides prior knowledge, and the spatial location helps differentiate the user from the background.
Furthermore, we integrate a LLM to enhance scenario recognition through logical reasoning. The knowledge derived from the LLM is subsequently transferred back to the local device to enable efficient, on-device inference.
Evaluated on both public and self-collected dataset, \systemname{} achieves effective multimodal fusion for both activity and scenraio recognition, outperforms the baseline by 12\% and 15\%, respectively.
\end{abstract}


\maketitle

\section{Introduction}
\label{sec:intro}
\begin{figure}
  \centering
    \includegraphics[width=1\linewidth]{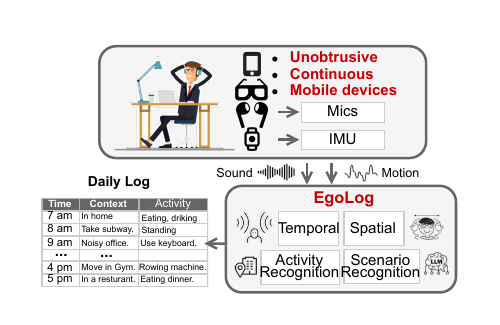}
    \vspace{-2em}
    \caption{\systemname{} leverages microphones and an IMU on commercial mobile devices for both activity and scenario sensing.}
    \label{fig:teaser}
    \vspace{-1em}
\end{figure}

Human activity recognition is a critical tool for monitoring physical and mental health, which enables numerous applications such as personalized healthcare and improving health outcomes \cite{alam2014gesmart, gedam2021review}.
These days, smartphones and other mobile devices are practically extensions of ourselves, capturing real-time data about our daily routines. 
For example, motion data from an Inertial Measurement Unit (IMU) can be used for human activity recognition since it directly correlates to the body movement. Various applications including  walking, standing, sitting, lying \cite{smartphonebased_recognition_of_human_activities_and_postural_transitions_341} are developed with IMU.
Besides, the built-in microphones can record audio data to support acoustic scene recognition \cite{martin2021low}, which is also relevant to human activity.

However, existing HAR methods on mobile devices are limited to coarse-grained, scenario-constrained detection. For instance, step counting is a common application, but it fails to capture diverse activities such as cycling, swimming, or strength training. Although prior work has extended sensing to vital signs, gait analysis, and exercise recognition, these approaches often focus on isolated tasks rather than continuous, holistic daily monitoring.

\new{This paper aims to design a mobile sensing system for ubiquitous wearables, such as earphones and smart glasses, to enable the continuous tracking of detailed egocentric human activities. Our approach involves modeling daily life from the user's perspective (i.e., an egocentric view), moving beyond the recognition of consecutive but isolated activities typically collected in laboratory settings. We observe two key distinctions between fine-grained daily logging and conventional activity recognition:
1. Daily life comprises long-term, continuous activity that contains richer information than short-term activity, which is called ``scenarios''. For instance, a user may spend an entire afternoon at the office, engaging in numerous interrelated activities.
2. Daily life inherently involves interactions with the environment. Consequently, sensors like microphones capture a complex auditory scene comprising sounds from the user, the environment, or a combination of both. While these environmental inputs can provide valuable contextual information to benefit activity recognition, they may also introduce significant noise.}

Several types of mobile devices are suitable for this task, including earphones, glasses, smartphones, and smartwatches, which users typically wear throughout the day and which possess rich sensing capabilities. The available sensors can be roughly divided into two categories: visual and non-visual.
Visual sensors, such as cameras, can capture a first-person view rich with information. However, they are less practical for daily logging due to several limitations: 1) a limited field of view (FOV), 2) privacy concerns, and 3) high energy consumption for continuous sensing. 
Instead, we envision incorporating non-visual sensors such as motion and audio sensors for our purpose. Those sensors are widely available on different mobile devices, including but not limited to smartphones, glasses, smartwatches, and earphones. As \fig\ref{fig:teaser} shows, \systemname{} can continuously monitor a user's daily activities in an egocentric manner and generate a detailed daily log, both the short-term activity and long-term scenario, with several examples illustrated in the figure.


However, there are some unique challenges below:
\begin{itemize}
    \item Existing mobile devices HAR models \cite{bhattacharya2022leveraging, gong2023mmg} lack the awareness of scenario information present in the real world. Specifically, user activities are driven by underlying reasons, and their distribution varies depending on the ``scenario.''  For instance, in an ``indoor cooking'' scenario, the ``jogging'' activity is less likely to happen.
    \item HAR models that depend on audio recordings are susceptible to interference due to the lack of spatial understanding. As a result, the audio data may lead to potential errors, which can be harmful for post-processing. For example, when detecting a ``clap'' activity based on sound, a nearby person's action could also trigger the model.
    \item Multi-modal fusion of audio and IMU is challenging and may not outperform the uni-modal model \cite{huang2022modality}. The heterogeneity of sensors indicates a weak correlation between them, where the effective fusion is particularly challenging and validated in MMG-Ego4D \cite{gong2023mmg}. Therefore, effectively extracting and leveraging sensor correlations is crucial for successful multi-modal fusion.
\end{itemize}

Towards resolving the above challenges, we propose the following designs of \textit{\systemname{}}:
First, we recognize the long-term scenario through temporal understanding, as detailed in Sec. \ref{sec:scenario}. This is achieved using a two-stage model: In the first stage, contrastive learning is employed to project audio and IMU features into a shared space, identifying well-aligned segments as informative key frames. The second stage then aggregates these key frame features, along with the original unimodal features, over time using a sequence model to achieve robust scenario recognition.

Secondly, we distinguish user-generated activities from environmental sounds (including those produced by other people) in Sec. \ref{sec:localization} by spatial understanding. Specifically, we formulate this as a sound localization task, where near-field sounds are treated as user-centric events, while far-field sounds are regarded as ambient background. To this end, we introduce a sound localization approach enhanced by natural human movement, which classifies sound sources into eight spatial regions. Our proposed motion transformer incorporates IMU readings to dynamically compensate for and refine localization estimates. This motion compensation module not only corrects errors induced by user motion but also alleviates common ambiguities in spatial perception, such as near-far confusion.

\new{
Third, we introduce an improved multi-modal network for activity recognition that integrates audio, IMU, and scenario information in Sec.~\ref{sec:har}. Unlike short-term sensor data (e.g., audio and IMU), the derived scenario captures long-term behavioral patterns and provides high-level semantic cues that significantly enhance recognition accuracy.
The network employs dedicated unimodal backbones for feature extraction and leverages a self-attention mechanism to dynamically fuse multimodal inputs, where the scenario information is encoded into a dense representation via a text encoder. 

Finally, in Section \ref{sec:llm}, we introduce a novel edge-cloud framework assisted by a Large Language Model (LLM)  to further enhance scenario recognition. This approach leverages the strong reasoning capabilities of LLMs to support a smaller-scale local model. Specifically, we convert multi-modal data, including audio, IMU signals, and the outputs from both sound localization and the local scenario recognition module, into contextual information for the LLM. When the local model's prediction confidence is low, the edge device queries the cloud-based LLM for inference. The refined scenario output generated by the cloud is subsequently downloaded to fine-tune the local model, thereby improving its performance over time.
}

We implemented and evaluated \systemname{} on three large-scale public dataset (e.g., Ego4D \cite{grauman2022ego4d}) and a self-collected dataset, which was collected by various form-factor from 10 participants, 12 activities for validation.
We demonstrate that \systemname{} obtains 12\% improvement on activity recognition and 15\% improvement on scenario recognition with low computation cost against strong multi-modal baseline (e.g., ImageBind \cite{girdhar2023imagebind}).
The contributions of this paper can be summarized as follows:

\begin{itemize}
\new{
    \item 
    We propose a new audio-IMU fusion method from two perspectives: temporal and spatial understanding. We extract scenario-level features through contrastive learning and aggregate them from short-term windows for scenario recognition. Besides, by integrating motion information with sound source localization, we boost the performance with movement compensation.
    
    \item 
    We propose a novel HAR method that takes both the scenario context and raw sensor data as input. On one hand, we distinguish similar activities by leveraging the extracted scenario information. On the other hand, we use the sound source and audio-IMU correlation to differentiate the user-generated sounds from background sounds.

    \item 
    We propose an LLM-assisted scenario recognition paradigm. In this framework, processed sensor data is fed into an LLM for high-level reasoning. The LLM's output is then used to transfer actionable knowledge back to the local device for on-device inference.
}     
    \item
    We evaluate \systemname{} on both the public dataset and the self-collected dataset, and conduct a user study on whether the generated summary can be helpful in daily life. Evaluations of \systemname{} on activity and scenario recognition demonstrate a performance improvement of up to 15\% over the best of our baselines.
\end{itemize}

\section{Related Work}

\label{sec:related}
\subsection{Motion Sensing for HAR}
IMU sensors in wearables and earables facilitate position tracking and user motion recognition, with their performance significantly impacted by sensor placement.
For example, LIMU-BERT \cite{xu2021limu} recognizes up to six activities with IMU recording of a smartphone.
Similarly, IMU on the wrist (e.g., smartwatch) can be used to recognize the basic activities \cite{chan2024capture}. 
Current motion sensing \cite{chan2024capture} only supports coarse-grained tasks, even with a large-scale dataset (>3000 hours), such as physical activity level (4 levels) or activities of daily life (10 activities).
Compared to smartphones and smartwatches, earables are a more reliable choice due to their relatively fixed position \cite{lyu2024earda, han2023headmon, han2023headsense, han2025headmon+}, as we illustrated before. Nonetheless, they share the same limitations in fine-grained motion sensing.

To enhance motion sensing, deploying multiple IMU sensors can enable precise full-body keypoint estimation with up to 17 IMUs. 
The pose of humans can be used to recognize fine-grained human activity, as demonstrated by MotionBERT ~\cite{zhu2023motionbert}. However, such an approach is impractical for everyday use. Instead, researchers propose leveraging commercial devices only, such as smartwatches, smartphones, and earables. For instance, combining multiple devices can enable fine-grained exercise activity recognition~\cite{stromback2020mm}. Similarly, IMUPoser achieves full-body pose estimation with just three IMUs~\cite{mollyn2023imuposer}. While using multiple devices improves performance, having all of them will have problems of high cost, user discomfort, setup complexity, and limited availability.

Another direction involves using pre-trained models to fuse IMU data with other modalities. Models like IMU2CLIP \cite{moon2023imu2clip} and ImageBind \cite{girdhar2023imagebind} link motion data to text or vision, enabling zero-shot recognition via retrieval. However, their performance on fine-grained tasks remains limited. Recently, Large Language Models (LLMs) have shown promise in reasoning over motion data \cite{ji2024hargpt, leng2024imugpt, xu2024autolife, ouyang2024llmsense}, often by converting sensor readings into text tokens for the LLM to process \cite{xu2024penetrative}.



In summary, prior work on motion sensing either employs multiple IMU sensors or remains confined to coarse-grained activity recognition. Since IMU sensing inherently
lacks contextual or environmental information, which limits their ability to perform fine-grained HAR \cite{tong2020accelerometers}. Further advancements should focus on integrating additional modalities or external knowledge.

\subsection{Acoustic Sensing for HAR}
In addition to IMU sensors, acoustic sensors (microphones and speakers) are also commonly integrated into wearables and mobile devices. The use of a speaker allows acoustic sensing to be categorized into two approaches: active sensing (utilizing both the speaker and the microphone) and passive sensing (using only the microphone).

As the main research area of audio, there are multiple applications of passive audio sensing, including speech recognition and audio classification \cite{schmid2023efficient}. However, audio is related but does not directly reflect human activity. Instead, it relies on the sound caused by activity, which can be easily misled by a loudspeaker. As a result, audio-based HAR is often treated as a specialized form of audio classification \cite{cristina2024audio}.
In contrast, multi-channel passive sensing enhances spatial awareness by estimating the direction of arrival (DoA) of sounds using microphone arrays \cite{schmidt1986multiple, adavanne2018sound} or binaural earables \cite{yang2022deepear, krause2021joint}. While DoA provides useful context, it cannot confirm whether the source is human or differentiate the loudspeaker. To detect human presence, some approaches integrate feedforward microphones to capture vocal activity via head-related sounds \cite{zhang2024earsavas}, or use an IMU to capture the bone-conduction sound \cite{he2023towards}. However, they can only cover the vocal activity.

Different from passive sensing, active sensing analyzes transmitted acoustic waves and their reflections, enabling non-contact sensing.
For the context of human activity, we are more interested in 
sensing not only the facial movements \cite{dong2024rehearsse} with outward-facing speaker, such as upper body pose detection \cite{mahmud2023posesonic}, and gesture recognition \cite{yang2024maf}. However, the speakers are usually oriented inward or designed to be effective only in the near field to maintain user privacy, which hugely limits the performance in coarse-grained recognition. 

In summary, acoustic sensors in wearables enable rich information capture through either passive or active sensing. However, audio-based HAR is still limited, as passive sensing struggles to confirm human presence, and active sensing is constrained by practical considerations, necessitating multi-modal integration for improved performance.

\subsection{Multi-modal Sensing for HAR}
Multi-modal learning of audio and IMU data addresses limitations in motion sensing or acoustic sensing by leveraging the complementary strengths of both modalities. For instance, VibVoice \cite{he2023towards} uses bone-conduction vibration and audio in earables to distinguish a user’s voice from others. Similarly, \cite{min2018exploring} employs IMU for motion detection and a microphone for conversation detection, processed independently. Other wearables, such as wrist-mounted or head-mounted devices, also utilize multi-modal approaches; SAMoSA \cite{mollyn2022samosa, bhattacharya2022leveraging} integrates audio and IMU on smartwatches to recognize up to 26 activities. However, multi-modal learning is not universally superior, as its effectiveness depends on factors like data quality, modality correlation, and model design. 
For example, the fusion of audio and IMU on a diverse dataset like Ego4D \cite{grauman2022ego4d} can be challenging.
As demonstrated by MMG-Ego4D \cite{gong2023mmg}, where a multi-modal model for smart glasses underperformed compared to single-modal models in recognizing nearly 200 actions, highlighting that multi-modal approaches do not always guarantee improved performance \cite{huang2022modality}.

Beyond integrating audio and IMU sensors, incorporating additional knowledge sources can enhance motion sensing. For example, SAMoSA \cite{mollyn2022samosa} uses automatic context detection to trigger context-specific classifiers, though limited to predefined classes. Furthermore, the advanced LLMs facilitate multi-modal processing, either at the prompt level, incorporating IMU, GPS, Wi-Fi, or Bluetooth data \cite{yang2025contextagent, xu2024autolife,yang2024drhouse}, or at the token level, combining audio and IMU \cite{han2023onellm, girdhar2023imagebind}. While promising, those works lack explicit multi-modal fusion, especially for the heterogeneous multi-modal data from mobile devices.

In summary, the fusion of audio and IMU can be challenging since the correlation between them can be weak. While incorporating external knowledge or using powerful models like LLMs for fusion can enhance recognition, current methods still lack effective techniques for deeply integrating the heterogeneous data.
\section{Motivation Study}
\label{sec:measurement}
As discussed in Sec. \ref{sec:intro}, we mainly have three challenges towards a multi-modal HAR on mobile devices: 1) lack of scenario awareness, 2) being sensitive to ambient noise, and 3) fusion collapse. To explore these challenges and evaluate the potential for overcoming them, we conduct a series of experiments to analyze and demonstrate promising directions.

\begin{figure}
    \centering
    \includegraphics[width=0.8\linewidth]{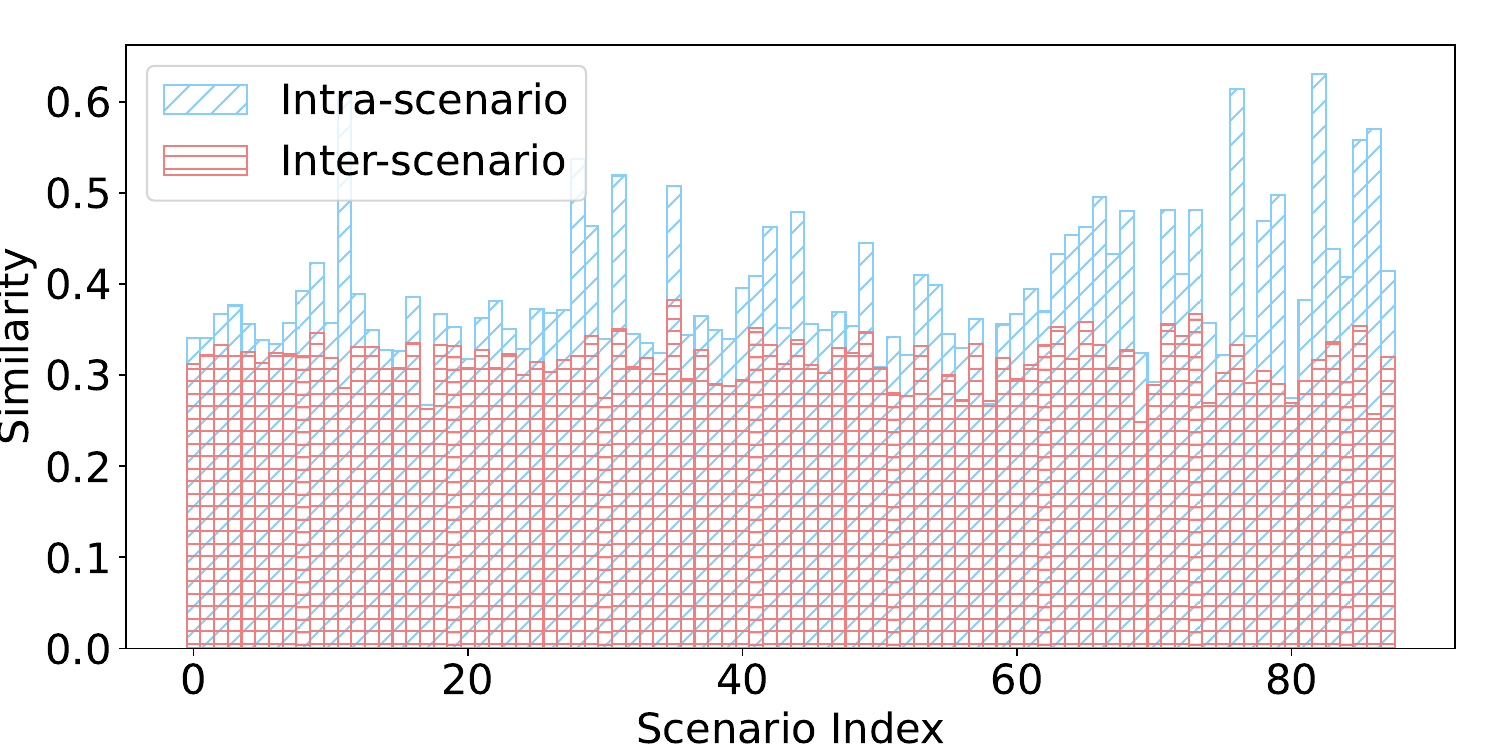}
    \vspace{-1em}
    \caption{Motivation for scenario understanding: 
    The similarity of embeddings for intra- and inter-scenario activities.}
    \label{fig:scenario_similarity}
    \vspace{-1em}
\end{figure}

\subsection{Scenario}
\label{subsec:scenario}
Scenario information refers to long-time activities (e.g., playing with a pet or working at home) that reflect the user's real-world behavior. Capturing such scenarios in datasets is challenging, as it requires volunteers to perform activities naturally, mimicking real-life conditions. Generally, scenario is the ensemble of multiple related and continuous activities with a realistic distribution, reflecting meaningful interactions with the environment.

To investigate the impact of scenario information on HAR, we utilize the Ego4D dataset \cite{grauman2022ego4d}, which captures real-world activities through smart glasses equipped with IMU and audio sensors. The dataset provides annotations at both the window level (every 10 seconds) and the video level, represented as plain text. We treat window-level annotations as activity labels and video-level annotations as scenario labels, reflecting the broader scenario of prolonged activities. To examine the properties of scenarios, we employ Sentence-BERT \cite{reimers2019sentence} to generate text embeddings for activities across 91 scenarios. We then compute activity similarity both within scenarios (intra-scenario) and between scenarios (inter-scenario). Intra-scenario similarity measures the similarity of activities within the same scenario, while inter-scenario similarity compares an activity from one scenario to a randomly selected activity from a different scenario.

The results are shown in Fig.~\ref{fig:scenario_similarity}, revealing that the similarity within a scenario (intra-scenario similarity) is over 20\% higher than the similarity between different scenarios (inter-scenario similarity). This suggests that each scenario has a distinct distribution of activities. Specifically, a scenario effectively represents the user's daily log since it encompasses a longer time frame.
Additionally, we notice that the inter-scenario similarity is generally above 0.3, indicating that many basic activities (such as sitting and walking) are common across all scenarios. Therefore, it is crucial to eliminate the influence of these basic activities in scenario recognition, as they are present in every scenario.

In summary, the study on the in-the-wild dataset indicates unique activity distributions per scenario, which can be leveraged to recognize scenarios by identifying scenario-specific activities.

\subsection{Sound Location }
\label{subsec:embodied}
\begin{figure}
    \centering
    \begin{minipage}{0.48\linewidth}
        \centering
       \includegraphics[width=\linewidth]{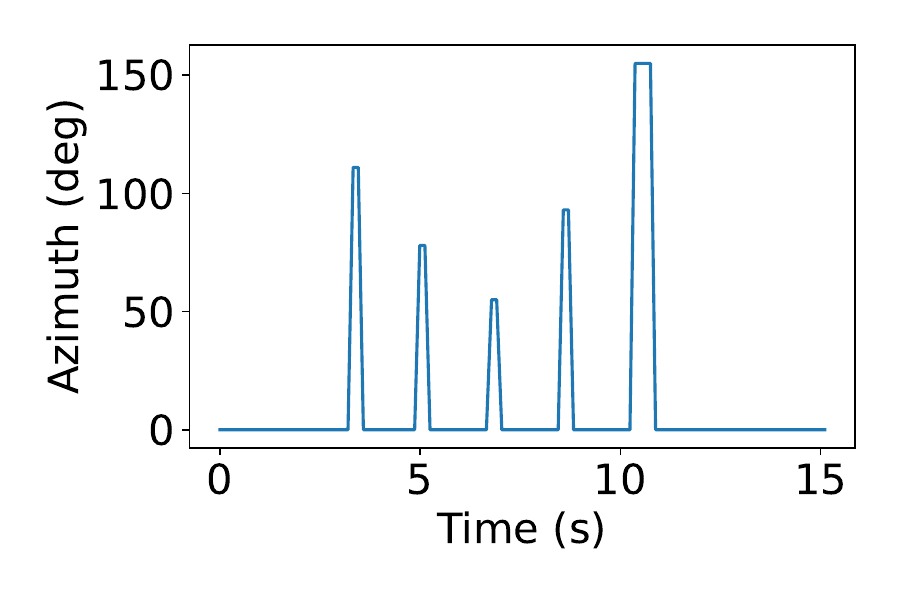}
        \caption{Sound localization \cite{schmidt1986multiple} under motion}
        \label{fig:loc_move}
    \end{minipage}
    \hfill
    \begin{minipage}{0.48\linewidth}
        \centering
        \includegraphics[width=\linewidth]{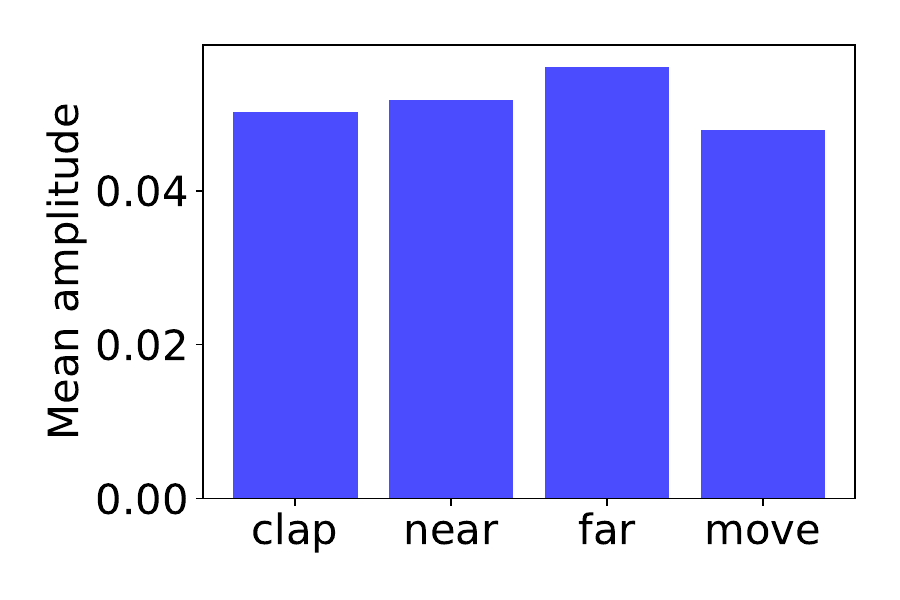}
        \caption{Sound volume may not related to distance.}
        \label{fig:amplitude}
    \end{minipage}    
    \vspace{-1em}
\end{figure}

\begin{figure}
    \centering
    \includegraphics[width=0.75\linewidth]{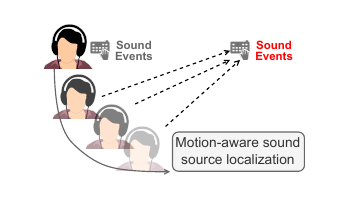}
    \vspace{-1em}
    \caption{Motivation for spatial understanding: estimate the distance of a sound event by motion.}
    \vspace{-1em}
    \label{fig:motivation_localization}
\end{figure}

In practice, recorded audio often includes far-field sound sources that are unrelated to the user’s activities, underscoring the importance of sound localization for effective HAR. Wearables that are consistently worn on the user’s head provide a stable and reliable position for capturing detailed ambient sound information. Additionally, modern wearables feature multi-channel recording capabilities, enhancing their spatial sensing potential.
In the following, we divide sound localization into direction and distance estimation, which we discuss as follows.

\new{
The estimation of sound direction is one of the mature, basic, but still challenging applications of a microphone array. The accuracy highly depends on the number of microphones, which is unfortunately not sufficient for most commercial devices. Considering the most common setting of stereo recording, there are mainly two problems: the deviation of direction (especially for front-back confusion) and estimation of sound distance, as illustrated in Fig. \ref{fig:motivation_localization}.
We envision that the above challenges may be solved by motion compensation. Specifically, we can aggregate the consecutive sound localization results along with the movement of the user (microphone) to refine the results or even get the distance.

To illustrate them, we positioned a sound source in front of the user and utilized a binaural microphone along with the algorithm from \cite{schmidt1986multiple} to estimate the direction of arrival, as shown in Fig. \ref{fig:loc_move}.  The user is instructed to move the body slightly and naturally. We found that the estimation was not satisfactory, as it fluctuated from the true value of 90 degrees. To validate whether sound volume is enough for differentiating near and far, we conducted an experiment where the sound source was at different distances from the microphone. As shown in Fig. \ref{fig:amplitude}, there was no notable difference in volume between different cases, indicating that volume alone is an unreliable indicator for distance. The final bar in Fig. \ref{fig:amplitude} reveals that user movement slightly affects amplitude. Moreover, the presence of multiple sound sources can result in an even more complex mix of random sound volumes.
}

\subsection{Multi-modal Fusion }
\label{subsec:measure-fuse}

\begin{figure}
    \centering
     \includegraphics[width=.75\linewidth]{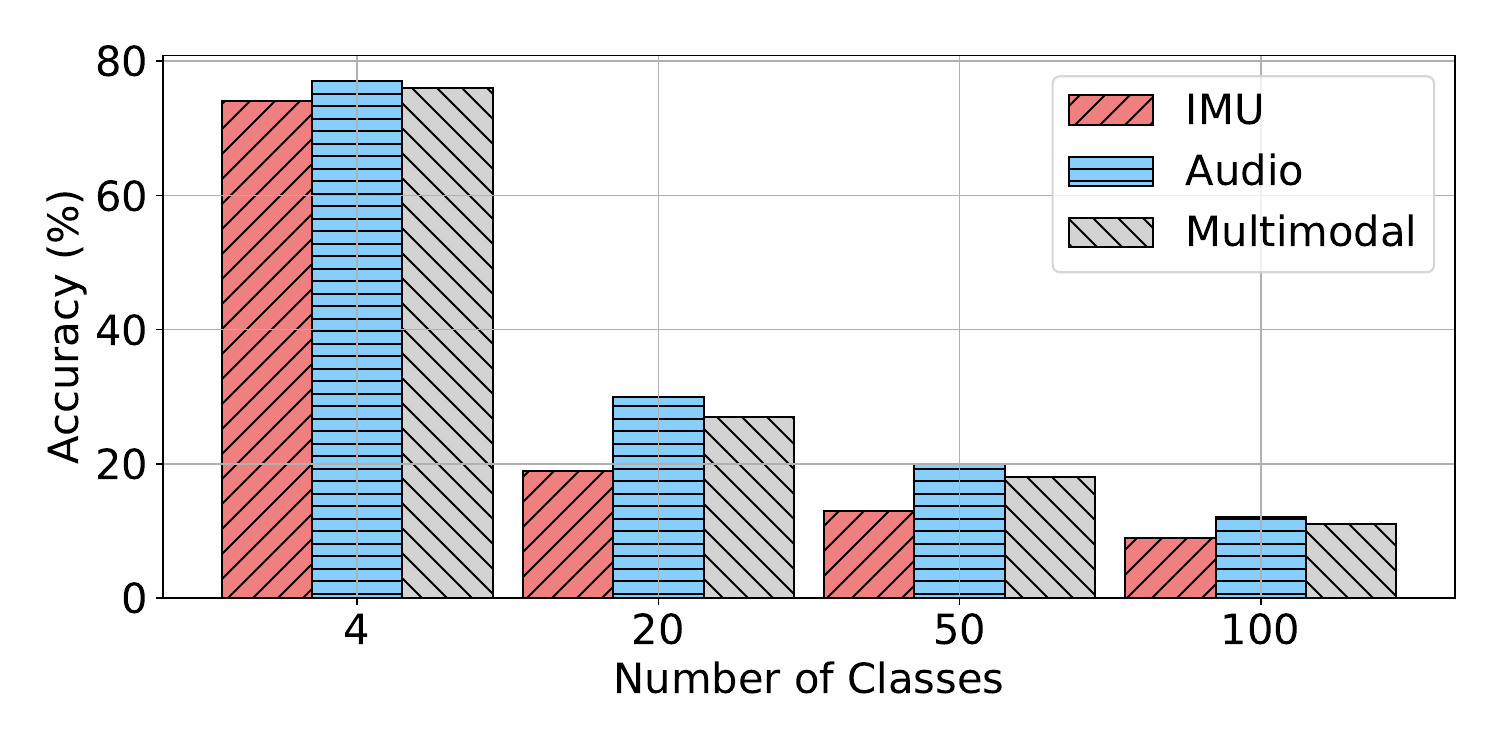}
     \vspace{-1em}
      \caption{The performance of an existing activity recognition with different class numbers and modalities.}
    \label{fig:num_class}
    \vspace{-1em}
\end{figure}

To investigate scalability bottlenecks in activity recognition using wearables, we trained unimodal supervised models with audio and IMU data on the Ego4D episodic memory subset (200 activity classes). We used EfficientAT~\cite{schmid2023efficient} for audio and LIMU-BERT~\cite{xu2021limu} for IMU as feature extraction backbones, comparing them to a multi-modal model via late fusion. Activities were clustered into varying class numbers using Sentence BERT~\cite{reimers2019sentence} embeddings and K-Means~\cite{macqueen1967some}, following MMG-Ego4D~\cite{gong2023mmg}. Results (Fig.~\ref{fig:num_class}) show a significant performance drop as class numbers increase, indicating challenges in recognizing diverse activities with audio or IMU alone. Multimodal fusion often underperforms unimodal models, suggesting ineffective extraction of complementary features at larger scales.

\begin{figure}
    \centering
    \includegraphics[width=.75\linewidth]{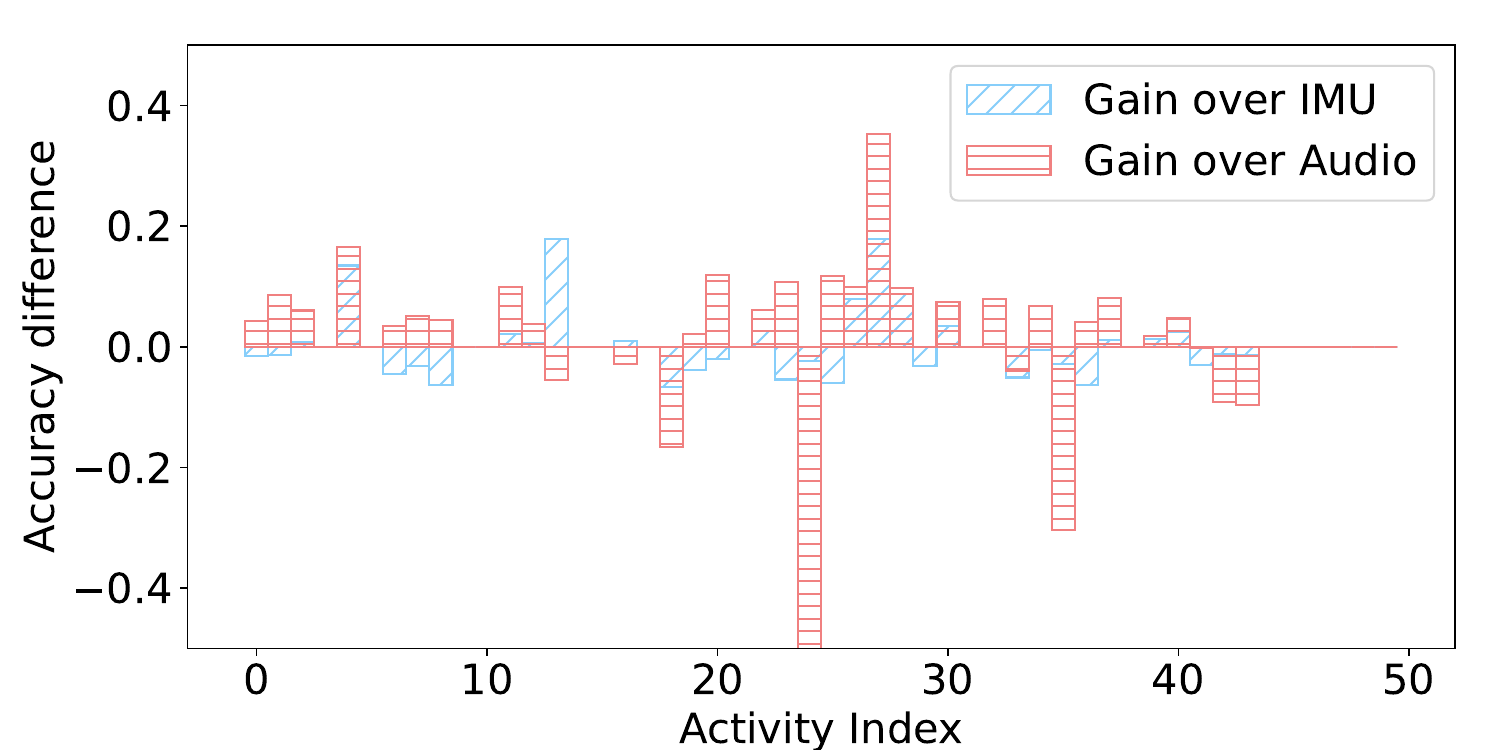}
    \vspace{-1em}
    \caption{Performance gain from multi-modal fusion.}
    \label{fig:acc_gap}
    \vspace{-1em}
\end{figure}

Previous work~\cite{mollyn2022samosa} demonstrates that multi-modal learning can outperform the unimodal model due to the complementary nature of each other.
However, it may not apply to all cases, where another study~\cite{gong2023mmg} finds multi-modal learning with audio and IMU on Ego4D leads to even worse performance than the single-modal solution. To investigate the reasons behind the unexpected results, we compared the accuracy of multi-modal and uni-modal models for each class, as shown in Fig. \ref{fig:acc_gap}. We observe that the advantages of multi-modal fusion fluctuate for both modalities. This indicates that supervised multi-modal learning can not automatically select the right modality, failing to benefit from modality fusion.

\new{
To investigate the effectiveness of LLMs in alleviating the problem, we employ the ImageBind-LLM \cite{han2023imagebind}, which converts both modalities into features and puts them in front of text embeddings. We instruct the model by the following prompt: ``<sensor> According to the audio and IMU reading on the earphone of the user, what is the activity of the user?'', while <sensor> refers to the place of sensor features. Since the LLM can have random output, we also add a prompt to force it to output one of the potential activities. To reduce complexity, we selected the top 30 activities in the subset of Ego4D (episodic memory). The BERT score of ImageBind-LLM is 0.04, 0.07, and 0.05 for precision, recall, and F1-score, respectively. As long as the performance is too low, we inspect the output and find that the LLM may not fully understand the task (multi-modal fusion). 
Furthermore, we inspect whether fine-tuning the model can resolve the problem by training an adapter~\cite{zhang2023llamaadapter} for ImageBind-LLM. The adapted LLM achieves precision, recall, and F1-score for BERT Score of 0.47, 0.46, and 0.47, respectively. Although it is improved a lot, it is still much lower than usability and is far from practical. 
}
\section{System Design}
\begin{figure*}
    \centering
    \includegraphics[width=0.8\linewidth]{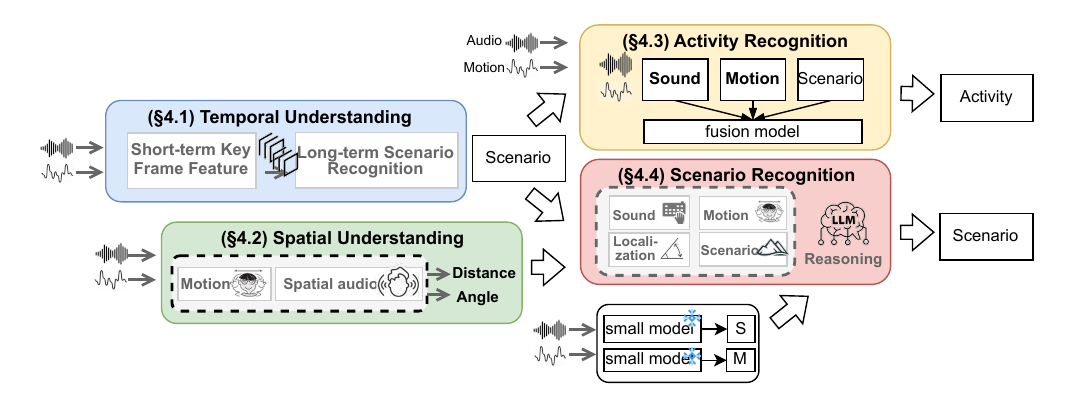}
    \vspace{-1em}
      \caption{The system overview of \systemname{}: the temporal and spatial understanding are two modules for multi-modal fusion, providing important input to the later modules to recognize activity and scenario}
      \vspace{-1em}
    \label{fig:system}
\end{figure*}

Figure~\ref{fig:system} provides an overview of \systemname{}, a mobile system designed to capture egocentric daily logs over both long-term and short-term periods. 
Specifically, Section~\ref{sec:scenario} and Section~\ref{sec:localization} detail the fusion of audio and IMU data to achieve a deeper understanding of scenario (i.e., long-term descriptions) and spatial information (i.e., the relative location of sound sources). These insights serve as foundational clues for subsequent processing. 
In Section~\ref{sec:har}, we integrate the scenario information into the multi-modal activity recognition process, effectively mitigating modality collapse. Finally, Section~\ref{sec:llm} describes the use of an LLM as an expert to refine the scenario recognition from the previous stage; the resulting knowledge is subsequently transferred back to the local device.

\subsection{Temporal Understanding}
\label{sec:scenario}

\begin{figure}
  \centering
    \includegraphics[width=1\linewidth]{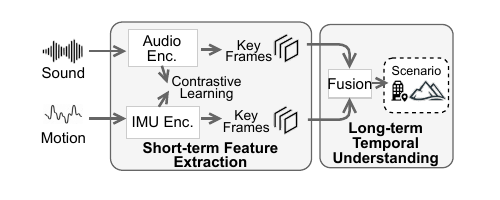}
    \vspace{-1em}
    \caption{Temporal understanding: we use contrastive learning to extract the cross-modal feature, which is important for the temporal-level fusion that recognize the scenario.}
    \vspace{-1em}
    \label{fig:scenario-recog}
\end{figure}

The results of our measurement study indicate that recognizing a large number of activities using IMU or audio data with an end-to-end model is challenging.
Therefore, we propose a scenario recognition approach to provide a prior information that can assist in activity recognition (as detailed in Section \ref{sec:har}). Additionally, scenario recognition itself is valuable for gaining a better understanding of daily logs.
As previously discussed, a scenario can be considered as a long-duration activity composed of multiple correlated sub-activities. To achieve accurate recognition, it is essential to develop an improved method for temporal analysis.
The pipeline is illustrated in Fig.~\ref{fig:scenario-recog}.

\subsubsection{Problem analysis}
The audio and IMU data represent their own characteristics and different advantages. As a result, the pretrained encoders tailored for these modalities are unique and fine-tuned for particular objectives. This variation suggests that simply merging the output from these encoders might not produce the optimal outcomes, as corroborated in Sec.~\ref{sec:measurement}. Thus, it is crucial to thoughtfully integrate these diverse modalities to fully leverage their collective advantages.

We aim to tackle scenario recognition, which refers to identifying high-level human activities over an extended period (over 30 seconds). As detailed in Sec.~\ref{sec:measurement}, current pre-trained encoders already hold some scenario-level information, although they still fall short of ideal performance. A straightforward approach involves training a classification model using a dataset annotated with scenarios in an end-to-end fashion. Nonetheless, we have identified two persisting challenges: (1) the end-to-end model may struggle to generalize across different input lengths. (2) The performance might be inferior if scenarios are not included in the training data. Therefore, we propose training a new encoder for two modalities to obtain scenario-level features from brief time windows, which will then be applied in a secondary scenario recognition phase.

Given audio data and IMU data, the corresponding annotations for activity and scenario are assigned. Ideally, we could estimate activities individually and recognize scenarios accordingly. However, many activities may not relate to the audio or IMU data, such as silent motions that exceed the audio's capability.
Instead, we propose using two encoders: one for audio and one for IMU data. Their outputs would project the same feature space, representing features of the same scenario. Like activities, the two modalities may not always directly correlate with the scenario, so the mapping can fail sometimes. To address this, we can estimate the scenario by combining multiple windows of data, compensating for the sparsity of features.

\subsubsection{Contrastive learning}
Based on the preceding analysis, employing contrastive learning is optimal for capturing scenario-level features. Contrastive learning focuses on acquiring low-dimensional representations ($SE_i$) by differentiating between similar and dissimilar data samples. The method aims to group similar samples closely in the representation space while distancing dissimilar ones using the Euclidean distance. The efficacy of contrastive learning has been demonstrated in earlier research~\cite{han2023onellm, girdhar2023imagebind}. 
Within our framework, similarity and dissimilarity can be defined in two ways. Firstly, any data point aside from itself is deemed dissimilar. Secondly, data points within the same scenario are viewed as similar, whereas others are regarded as dissimilar. We expect that the first approach will yield more robust features, since it does not rely on explicit scenario annotations, and this will be assessed during the evaluation.

Regarding the network design, we employ two distinct deep neural networks to derive embeddings: LIMU-BERT~\cite{xu2021limu} for the IMU data and EfficientAT~\cite{lyu2024earda} for the audio input. Each sample spans a two-second window. The training of both encoders is based on the cross-entropy loss function, structured as follows:
\[
logits = (Audio_e \cdot IMU_e^T) \cdot \exp(\tau),
\]
where $Audio_e$ and $IMU_e$ refer to the features from two encoders, and $\tau$ refers to the learned temperature parameter \cite{chen2020simple}. Then, we get the loss using cross-entropy loss as follows:
\[
L = cross\_entropy\_loss(logits, label, axis=0).
\]

\subsubsection{Scenario recognition}
\new{
Assuming we extract the features from $Audio_e$ and $IMU_e$ by contrastive learning, they may not immediately convey the scenario as previously noted. We observe that a robust example in which both modalities relate to the scenario will exhibit a strong similarity between $Audio_e$ and $IMU_e$. This high similarity indicates that both modalities are effectively mapped into the same feature space, highlighting their importance as key features.
Conversely, low similarity suggests that the two modalities have not been successfully integrated into a unified representation of the scenario. Ideally, we would only keep frames with high similarity. However, the dynamics of execution can be problematic, introducing instability. Furthermore, it is difficult to set a definitive threshold to classify all frames below it as "useless."
Given the inherently weak correlation between audio and IMU, examples with high similarity may be rare. Therefore, it is important to leverage as much information as possible from the available data.
To address this problem, we propose extending the key features along the temporal dimension. We integrate a sequence model—either an LSTM or a transformer—to compile features over time (as shown in the right section of Fig.~\ref{fig:scenario-recog}).
This two-stage architecture enables the model to capture both short-term and long-term dependencies within the features, thereby improving the accurate recognition of extended scenarios. Finally, a linear layer followed by a sigmoid activation function is used to predict the probability for each scenario.
}

\begin{figure}
  \centering
    \includegraphics[width=1\linewidth]{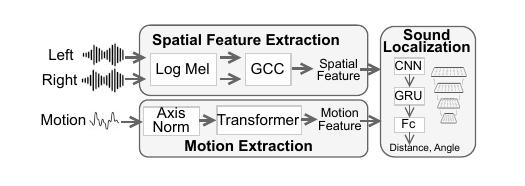}
    \vspace{-1em}
    \caption{Spatial understanding: we differentiate sound by location by the combination of spatial audio and motion, leveraging the natural but random movement of the user.}
    \label{fig:seldnet}
    \vspace{-1em}
\end{figure}

\subsection{Spatial Understanding}
\label{sec:localization}
Existing activity recognition methods by audio are often misled by environmental events.
As illustrated in Sec. \ref{sec:measurement}, it is not enough to use sound volume to differentiate the near-field and far-field sound. Here, we are interested in the near-field sound because of the close correlation between it and the user's interaction with the environment. 

\subsubsection{Sound localization}
The location of sound can be estimated by a multi-channel microphone array. 
Specifically, the time of arrival of the sound source for one microphone is: \(t_i = \frac{\sqrt{(x_s - x_i)^2 + (y_s - y_i)^2}}{c}\), where the location of microphones $\mathbf{p}_i = (x_i, y_i)$, $i = 1, \ldots, N$, and a sound source at $\mathbf{s} = (x_s, y_s)$.
Since we are not aware of the start time of the sound source, we can only obtain the Time Difference of Arrival (TDoA) between two microphones like $i$ and $j$ instead. Considering there are only two microphones for binaural earables, we can only have one equation that is less accurate and can't solve front-back confusion.

Building on the work of SELDNet \cite{adavanne2018sound} and DeepEar \cite{yang2022deepear}, we propose a neural network-based binaural localization. Overall, our model performs an 8-class classification, distinguishing between four directions (front, left, right, back) and two distance categories (near and far). The binaural audio input, is converted to two log-Mel spectrograms and generalized cross-correlation coefficients of spectrograms, combined along the channel dimension to form spatial features with shape \( B \times 3 \times T \times 64 \), where \( B \) is the batch size and \( T \) is the number of frames. We process the audio feature with three convolutional layers, transforming it to (B, 128, T/5, 64). Then it is followed by a GRU layer, multi-head attention (resulting in B, 128, T/5, 64), and a linear layer that outputs (B, T/5, C).

\subsubsection{Movement compensation}

\new{
As illustrated in the motivation study, user movement can be leveraged to enhance spatial understanding. While similar concepts have been implemented in prior works such as \cite{zhu2021localizing, seo2022ross}, these approaches require either explicit user gestures or a fully controllable robot. In contrast, our proposed \systemname{} operates without relying on predefined gestures or manual intervention, as it utilizes natural, unintentional body movements that occur in daily life.

To realize this, we introduce an end-to-end method where IMU data—structured as (B, 6, T) and comprising both accelerometer and gyroscope measurements—is processed through two transformer layers. This module outputs a representation of shape (B, 384, T), which is then passed through a linear layer to produce an output of size (B, T/5, C). The resulting features are subsequently concatenated with the output from the GRU module.
The overall model performs an 8-class classification (C=8), distinguishing between four spatial directions (front, left, right, back) and two distance categories (near and far).

In the scope of \systemname{}, we can not assume that only one sound source exists. Instead, we may encounter the case where there are multiple sound sources overlapping. Since both sound localization and audio recognition can be multi-label, the problem becomes an association confusion, and we will discuss how to leverage the information in a later section.
}

\subsection{Scenario-aware Activity Recognition}
\label{sec:har}

\begin{figure}
    \centering
    \includegraphics[width=1\linewidth]{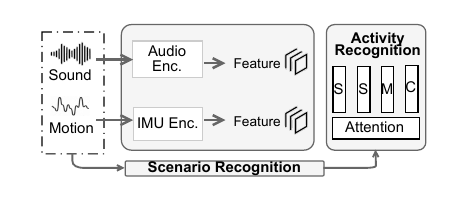}
    \vspace{-1em}
    \caption{Multi-modal activity recognition: We extract features from each individual modality (uni-modal features), combine them with the scenario information derived from temporal understanding, and merge them at the token level for final activity recognition.}
    \vspace{-1em}
    \label{fig:activity_recognition}
\end{figure}

The methods outlined in Sec. \ref{sec:scenario} and \ref{sec:localization} offer a promising path for audio-IMU fusion. Next, we propose to integrate them into HAR.

Our proposed multi-modal network consists of two main components: unimodal backbones and a fusion module. The unimodal backbones include separate feature extractors that gather features from each input modality. The fusion module is designed to combine and aggregate these features from the unimodal backbones, resulting in a fused output that denotes the activity of the user. Although the design described above shows promise, it shares similar input and output characteristics with those presented in Section \ref{sec:measurement}, which has already been validated. As a result, it may not perform well in human activity recognition due to a lack of multi-modal correlation.

There are two common approaches for fusing modalities: either use a multi-layer perceptron to process the concatenated representations of different modalities, or utilize a Transformer-based fusion module that processes a series of tokens from various modalities \cite{gabeur2020multi}. We have chosen the transformer-based fusion design because it can easily scale to accommodate any number of input tokens through attention mechanisms. 
This flexibility is particularly important, as our multi-modal model needs to handle data with a varying number of modalities (more than only audio and IMU) for the task. Specifically, suppose the scenario information is a feature with the same dimension as other modalities, the transformer module can easily integrate it.
The final output of the fusion module is calculated by averaging the output tokens, rather than relying on the CLS token. The formulation of the token for one modality $Z_m$ is below:
\[
Z_m = (Encoder_m(S_m), e_m), \quad Z=f(Z_1, Z2, Z_3...)
\], where \( m \) represents each modality, which includes audio, IMU, and text. The term \( e_m \) denotes the learnable embedding specific to each modality. The transformer fusion module will combine multiple \( Z_m \) tokens and produce a unified \( Z \) feature with consistent dimensions. 
Similar to the previous section, we keep the audio and IMU encoder used for scenario recognition since both of them are lightweight and can be deployed on a mobile device.
By default, we use the standard transformer with a number of layers of 2. 

\new{
Next, we extend the multi-modal model that can handle not only the audio and IMU, but also other auxiliary information, such as scenario:
\begin{itemize}
    \item Specifically, we collect scenario information from Sec. \ref{sec:scenario} in textual form. To integrate it, we use the ImageBind \cite{girdhar2023imagebind} text encoder to map this information into the same feature space as the other modalities. Note that the text embeddings can be precomputed and stored to reduce computational overhead for each scenario. When multiple scenarios are detected, we concatenate their corresponding text representations.

    \item For spatial information from Sec. \ref{sec:localization}, we treat it as an indicator for interference detection. When far-field sound is detected with a confidence score exceeding 0.9, the audio recording is considered unreliable. In such cases, we can either discard the audio recording and rely solely on motion sensing or apply distance-based sound extraction \cite{chen2024hearable} to retain only the near-field sound.
\end{itemize} 
}

\subsection{LLM-assisted Scenario Recognition}
\label{sec:llm}
\begin{figure}
    \centering
\includegraphics[width=1\linewidth]{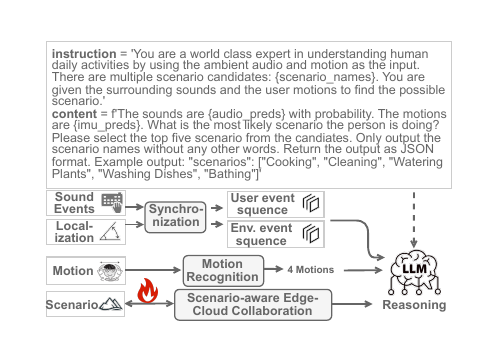}
\vspace{-2em}
    \caption{LLM-assisted scenario recognition: Utilizing prior outputs, we differentiate sound events into those originating from the user and those from the environment. We then combine it with motion and scenario recognition results and feed it into an LLM for reasoning (the upstream process). The LLM's output is then used as a pseudo-label to re-train the local model (the downstream process).}
    \vspace{-1em}
    \label{fig:llm_reasoning}
\end{figure}

As discussed in Section \ref{sec:measurement}, LLMs may be suboptimal for direct activity recognition. However, we posit they are well-suited for higher-level scenario recognition, as longer data sequences provide a richer information base. This aligns with the findings in Section \ref{sec:scenario}, where longer sequences also benefit multi-modal fusion—a property we expect to transfer effectively to LLM reasoning. As illustrated in Fig. \ref{fig:llm_reasoning}, our approach involves formatting sensor data (five seconds window) into text and using tailored prompts to guide an LLM in scenario recognition. The LLM's output can further refine local scenario predictions. It is important to note that all LLM processing is performed on the cloud, ensuring that \systemname{} remains a lightweight mobile system.
 
\subsubsection{Prompt design}
\new{
Existing studies primarily leverage the reasoning capabilities of LLMs on text-like data. This subsection introduces the fusion of multi-modal signals from audio and motion sensors for LLM-based reasoning in \systemname{}, as illustrated in Fig. \ref{fig:llm_reasoning}. Instead of re-training the LLM or adding an adapter, we focus on designing a specialized prompt to inject these multi-modal contexts. Specifically, we construct a demonstration within the prompt to guide the LLM in reasoning with such sensor data.
The prompt is structured as follows: “You are an expert in human activity analysis. You will receive audio and IMU recognition results, including extra information on potential scenarios. Using this information, reason and refine the person’s scenario. The scenario must be one of the following categories: [List of categories]. Your output must be a single line containing just one word or phrase.”
This prompt, along with the formatted input data as shown below, is fed to the LLM.

\para{Audio.} 
We employ a pre-trained audio recognizer \cite{schmid2023efficient} to transcribe audio into text by selecting the top three predictions. The output is structured as "the detected audio classes are <class1, prob1>, <class2, prob2>, <class3, prob3>...", where each entry is referred to as a sound event. To minimize noise, we retain only the top five predictions with a confidence score exceeding 0.3.
Since the classifier's label set is based on the AudioSet ontology, the defined classes are highly fine-grained, often resulting in multiple similar classes with comparable probabilities. This lack of discriminative power hinders effective reasoning. To address this, we apply a two-level hierarchical clustering of the classes to their parent categories, reducing the total number of classes from 416 to 50 and improving generalization.

\para{IMU.} 
Following the IMUCLIP approach \cite{moon2023imu2clip}, we classify IMU data into four activities: walking, moving, standing up, and sitting down. While previous works have demonstrated the ability to recognize more fine-grained activities using IMU data, their generalization capabilities remain questionable due to reliance on scenario-constrained datasets.

\para{Spatial.}
We leverage spatial information from Sec. \ref{sec:localization} by applying distinct processing strategies to near-field and far-field audio inputs, guided by tailored prompts. For near-field sounds, we use the prompt: “the <sound event> is happening in the near front of the user, which is likely to be related to human activity.” For far-field sounds, the prompt is: “the <sound event> is happening in the <sound location>.”
When both near-field and far-field sounds are detected, it is unclear which event is near or far. As a result, we can turn on the distance-based sound separation \cite{chen2024hearable} that seperate the audio into two and perform the sound recognition, respectively. We leave the optimization of the separation in the future work since it is unrelated to the main body of \systemname{}.
}

\subsubsection{Scenario-aware LLM Assistance}

\new{
We define an upstream and downstream collaboration paradigm where the edge device seeks assistance from a cloud-based LLM, rather than naively querying the response.

For the upstream path, we leverage the superior scenario recognition capabilities of LLMs, especially for out-of-domain data. The scenario estimated by the edge device can be treated as a noisy reference, providing valuable preliminary information for the LLM. We integrate this estimation, along with its confidence score, into the prompt supplied to the LLM. 
Specifically, the estimated scenarios are added to the prompt, augmented by the text: "the preliminary scenario estimation is <scenario, confidence>."
However, frequent queries to the cloud service raise concerns about cost, privacy, and latency. Ideally, all modules of \systemname{} should be executable on a mobile device if the performance is sufficient. Consequently, we propose a collaborative framework that strategically uses the cloud LLM to enhance the local model. Specifically, a confidence threshold (0.5) is set for the local scenario recognition model. A cloud query is triggered only when the local model's confidence is low.

Complementing the upstream process, we also propose a downstream collaboration where the cloud LLM's output can conversely assist the local model. Specifically, the data points associated with uncertain cases (which were uploaded to the LLM), along with the LLM's refined results, are cached and used to fine-tune the local model, which enables continuous on-device improvement.
Within the scope of \systemname{}, we do not consider adding new class to the local model since the required data can be huge, which leads to a long time for waiting.
}

\section{Evaluation}
\label{sec:evaluation}
\subsection{Experiment Setup}

\subsubsection{Implementation}
\begin{figure}[h]
    \centering
    \includegraphics[width=1\linewidth]{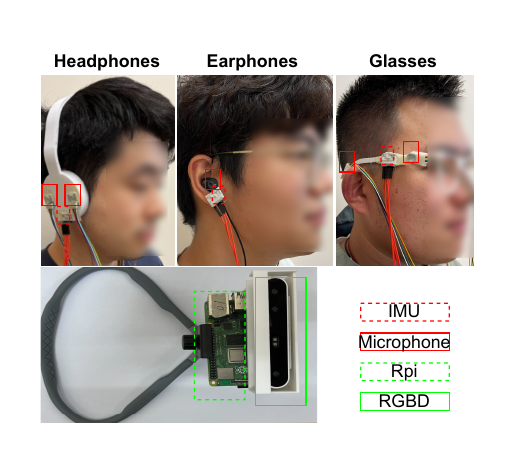}
    \vspace{-2em}
    \caption{Devices used in our data collection.}
    \vspace{-1em}
    \label{fig:device}
\end{figure}

\new{
We implemented our system on a Raspberry Pi 4, which was connected to multiple devices: binaural earables (MU6 \cite{Mu6_Home_Page_2020} and Sound Professionals \cite{soundprofessionals}), headphones, and smartglasses. These devices support multi-channel audio recording. Due to the lack of a commercial open platform with the same functionality, we prototyped our system using a 3D-printed model and a configurable microphone array, where the headphone prototype support 4-channel recording and the smart glasses support 4-channel, all devices are shown in Fig. \ref{fig:device}.
An IMU with a sampling rate of 200 Hz was attached to both the earables and the smartglasses.
The neural network was trained on a server equipped with an Intel i7-11700k CPU and two Nvidia RTX 3090 GPUs using PyTorch. For inference, all computational tasks—except for summary generation—are performed on the smartphone.
}

\subsubsection{Public dataset}
To comprehensively evaluate \systemname{} performance, we construct our dataset with the blow public datasets:

\para{Ego4D:}
\new{
Ego4D \cite{grauman2022ego4d} is a large-scale dataset collected by various brands of head-mounted cameras (e.g., GoPro), offering synchronized multi-modal data (i.e., video, audio, and IMU signal). We consider the moment subset of the dataset, which consists of 158 classes. We can perform k-means clustering on the text embeddings to adapt to different levels of granularity. For the scenario annotation, we use the scenario annotation \cite{grauman2022ego4d} for each video since they have a similar definition. In total, there are 91 scenarios in the dataset. 
Considering Ego4D is the most diverse and large-scale dataset, we set it as the default dataset.
}

\para{EgoADL:}
\new{
EgoADL \cite{sun2024multimodal} is also a large-scale dataset collected in the real world with Android smartphones, including Wi-Fi, audio, and IMU data. In this paper, we exclude Wi-Fi data for fair comparison. We perform a similar clustering for the 221 classes of actions in the dataset.}

\para{SAMOSA:}
\new{
SAMOSA \cite{mollyn2022samosa} is a relatively small-scale dataset collected by an Android smartwatch (i.e., Fossil Gen 5), including audio and IMU. There are 26 activities in total.
}

\subsubsection{Self-collected dataset}
\new{
The existing public datasets already include data from various devices—such as smart glasses, smartphones, and smartwatches—offering considerable duration and diversity. The purpose of our self-collected dataset is not to surpass these strengths but to address their limitations. Specifically, the spatial understanding discussed in Section \ref{sec:localization} relies on spatial audio, which is absent from the aforementioned datasets but is accessible using commercial devices. To compensate for this gap, we collected a new dataset featuring spatial audio recordings from different form factors: glasses, headphones, and earphones.

We constructed a dataset in which a volunteer acted as the user and a loudspeaker served as an audio interference source. The loudspeaker played audio samples from the NIGENS dataset \cite{trowitzsch2019nigens}, which contains 14 types of common sound events. The user, in contrast, performed the following 12 activities: standing up, sitting down, clapping, walking, talking, drinking, typing on a keyboard, eating, watching TV, washing hands, cleaning a room, and writing.
During data collection, the loudspeaker's position remained fixed. To simplify the experiment, the user's location was controlled for most activities (excluding walking), allowing us to manually record the relative position between the user and the loudspeaker. However, as users were not required to remain perfectly still, their precise position was not controlled with high accuracy. To ensure the dataset's validity, users were instructed to limit their positional variance to within the spatial resolution of our classification model. In essence, while the user's location was dynamic, the annotation was treated as static from the perspective of the classification task.
For the walking activity, where the relative position changes continuously, we used an RGB-D camera to perform Simultaneous Localization and Mapping (SLAM) to estimate the user's trajectory. This estimation was based on the assumption that the user started from a fixed, known position in the room (the location of the loudspeaker is also fixed and known).
}

\subsubsection{Baselines}
We adopt the following evaluation metrics in this paper. For the multi-label classification (i.e., scenario and spatial), we use the multi-label F1-Score: $2*TP/(2*TP+FP+FN)$. For the single-label classification (i.e., activity), we use the accuracy as the metric.

To evaluate our approach, \systemname{}, we have selected different baselines for activity recognition and scenario recognition, respectively. Specifically, we consider SAMoSA \cite{mollyn2022samosa} and Imagebind \cite{girdhar2023imagebind} as our baselines for activity recognition, which are conventional supervised multi-modal model that accepts both audio and IMU. For the scenario recognition, both of the above baselines with larger data duration are considered. To ensure a fair comparison, all of them are fine-tuned with the same data.

\subsection{Activity Recognition}
\begin{figure}[h]
    \centering
    \begin{minipage}{0.48\linewidth}
        \centering
      \includegraphics[width=1\linewidth]{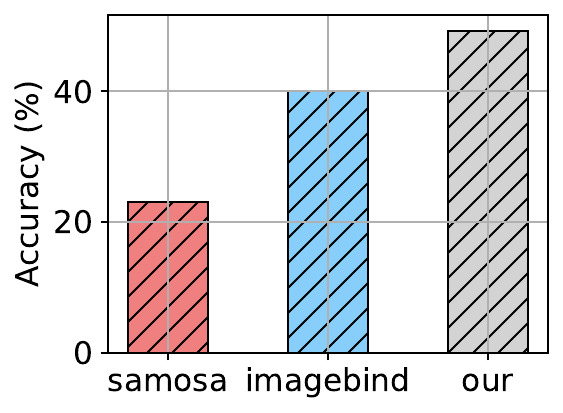}
          \vspace{-0.5em}
      \caption{Activity recognition on Ego4D.}
        \label{fig:activity_ego4d}
    \end{minipage}
    \hfill
    \begin{minipage}{0.48\linewidth}
        \centering
    \includegraphics[width=1\linewidth]{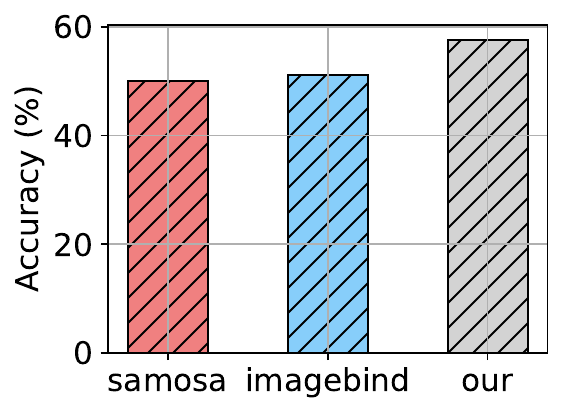}
        \vspace{-0.5em}
      \caption{Activity recognition on EgoADL.}
        \label{fig:activity_egoadl}
    \end{minipage}
    \vspace{-0.5em}
\end{figure}

\begin{figure}[h]
    \centering
    \begin{minipage}{0.48\linewidth}
        \centering
       \includegraphics[width=1\linewidth]{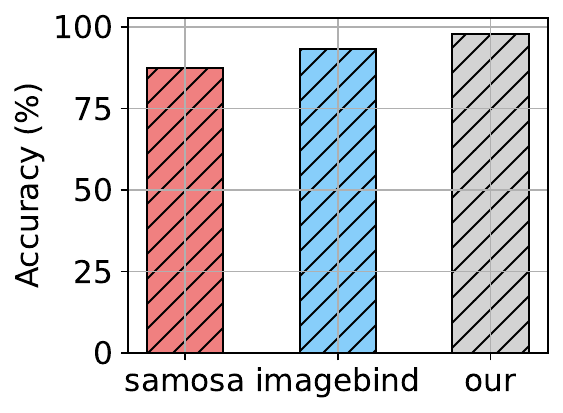}
           \vspace{-0.5em}
      \caption{Activity recognition on SAMOSA.}
        \label{fig:activity_samosa}
    \end{minipage}
    \hfill
    \begin{minipage}{0.48\linewidth}
        \centering
    \includegraphics[width=1\linewidth]{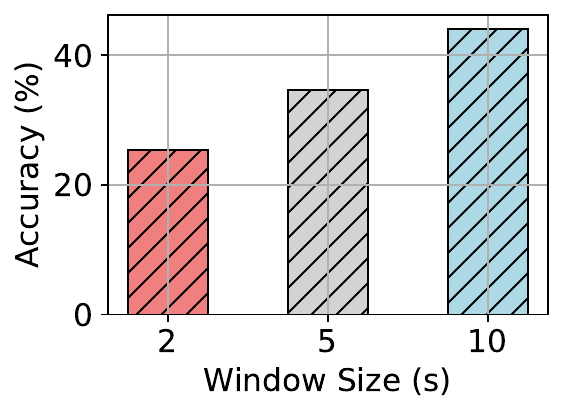}
        \vspace{-0.5em}
      \caption{Activity recognition vs. data length.}
        \label{fig:activity_length}
    \end{minipage}
\vspace{-0.5em}
\end{figure}

We present the performance of our method for human activity recognition on the Ego4D, EgoADL, and SAMOSA datasets. As shown in Fig. \ref{fig:activity_ego4d}, \ref{fig:activity_egoadl}, and \ref{fig:activity_samosa}, our approach consistently outperforms the baselines, with the most notable improvement on the SAMOSA dataset.

The performance gap is most significant on the Ego4D dataset (23\% vs. 52\%), which is the most challenging. Compared to ImageBind, \systemname{} achieves superior performance (52\% vs. 40\%) with significantly lower computational cost.

On the other two, less diverse datasets, the advantages of \systemname{} are slightly reduced. We attribute this primarily to a lack of contextual information; Ego4D data was collected "in the wild," while the others were gathered in moderately controlled experimental settings. Despite this, our approach still outperforms the best baseline by 6\%on EgoADL and 4\%on SAMOSA, demonstrating its robustness. Note that the absolute accuracy for SAMOSA is the highest, which is expected as it has the smallest number of classes, making it the least challenging dataset.

\para{Data length.}
Since the length of the data significantly influences both audio and IMU data, we first establish the hyperparameters before evaluating the system.
We present the performance of human activity recognition along with various data lengths, in Fig. \ref{fig:activity_length}. We observe that the HAR performance stays stable when the data length is larger than five seconds, while significantly dropping when the data length is one second. As a result, we keep the data length of HAR to five seconds. 

\begin{figure}[h]
    \centering
    \begin{minipage}{0.48\linewidth}
        \centering
       \includegraphics[width=1\linewidth]{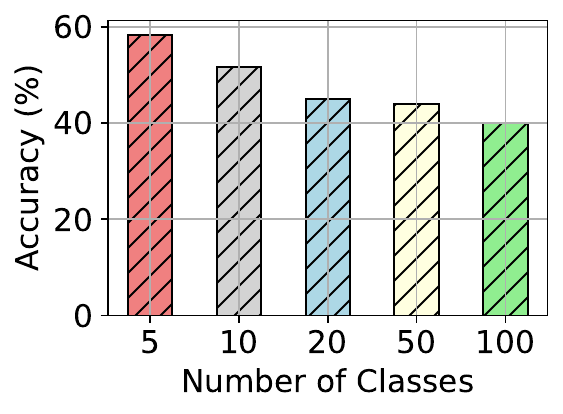}
           \vspace{-0.5em}
      \caption{Human activities recognition performance vs. number of classes.}
        \label{fig:activity_num_classes}
    \end{minipage}
    \hfill
    \begin{minipage}{0.48\linewidth}
        \centering
    \includegraphics[width=1\linewidth]{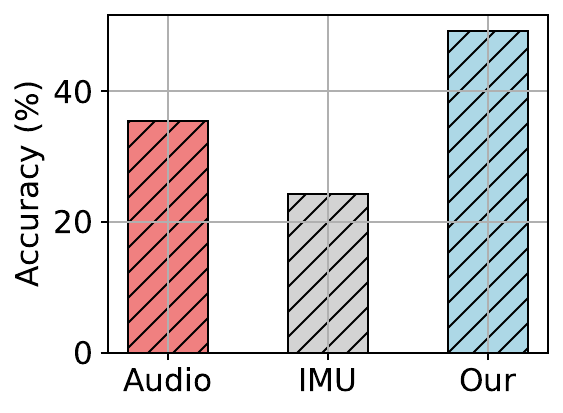}
        \vspace{-0.5em}
      \caption{Human activities recognition performance vs. modality.}
        \label{fig:activity_modality}
    \end{minipage}
\vspace{-0.5em}
\end{figure}

\para{Number of classes.}
We split the sampled Ego4D dataset into activities and evaluated their performance in Fig. \ref{fig:activity_num_classes}. We observe that there are performance gaps between activities, indicating that the complexity of each of them is different. Specifically, we inspect the top three worst activities, which are putting something in a container or on a surface, cutting with scissors, and sleeping. We observe that they are both quiet, and the motion is weakly related to the head, so it is natural to have lower performance.

\para{Modalities fusion.}
As shown in Figure \ref{fig:activity_modality}, our multi-modal fusion approach (audio+IMU) surpasses the audio-only and IMU-only baselines, improving activity recognition performance by more than 10\%.

\begin{table}[h]
\centering
\caption{Ablation study on activity recognition for different datasets, the metric is accuracy.}
\label{tab:ab_activity}
\begin{tabular}{ccccc}
\toprule
\textbf{Model}  & \textbf{Ego4D$\uparrow$} & \textbf{EgoADL$\uparrow$} & \textbf{SAMOSA$\uparrow$} \\
\midrule
Random  & 0.02 & 0.02 & 0.04 \\
\midrule
ImageBind-FT  & 0.40 & 0.51 & 0.93 \\
\midrule
Multi-modal & 0.401 & 0.49 & 0.90\\
+ Scenario & \textbf{0.467} & \textbf{0.58} & \textbf{0.97} \\
\bottomrule
\end{tabular}
\end{table}

\para{Ablation study.}
We conduct an ablation study on different components of our system, including each modality (audio and IMU), multi-modal fusion, scenario as condition, and scenario as condition via Imagebind-embedding, as shown in Tab. \ref{tab:ab_activity}. For reference, we also list the performance of a random guess (indicating the number of classes) and baseline ImageBind-FT.
We observe that

\subsection{Scenario Recognition}
\begin{figure}[h]
    \centering
    \begin{minipage}{0.48\linewidth}
        \centering
    \includegraphics[width=1\linewidth]{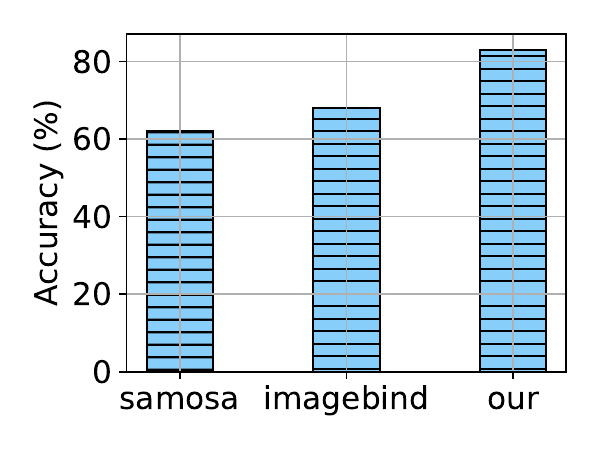}
           \vspace{-0.5em}
      \caption{Scenario recognition accuracy on Ego4D.}
        \label{fig:scenario}
    \end{minipage}
    \hfill
    \begin{minipage}{0.48\linewidth}
        \centering
    \includegraphics[width=1\linewidth]{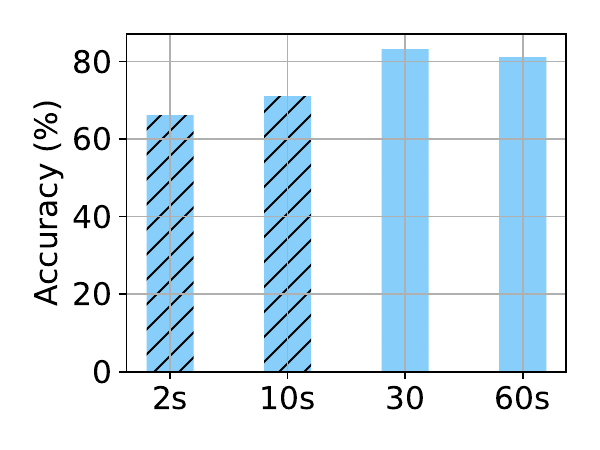}
        \vspace{-0.5em}
      \caption{Scenario recognition accuracy vs. data length.}
        \label{fig:scenario_length}
    \end{minipage}
    \vspace{-0.5em}
\end{figure}

\begin{figure}[h]
    \centering
    \begin{minipage}{0.48\linewidth}
        \centering
      \includegraphics[width=\linewidth]{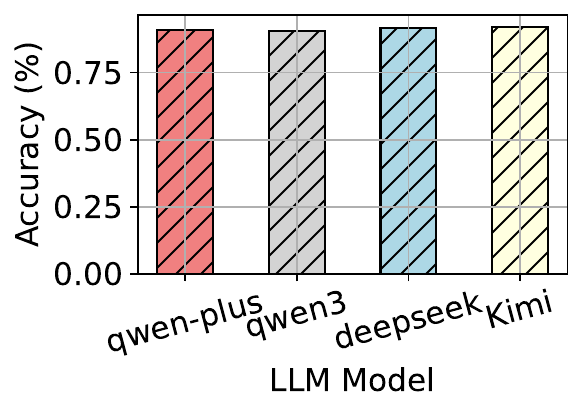}
          \vspace{-0.5em}
     \caption{Selection of LLM.}
    \label{fig:selection_llm}
    \end{minipage}
    \hfill
    \begin{minipage}{0.48\linewidth}
     \begin{tabular}{cc}
        \toprule
        \textbf{Model} & \textbf{Ego4D$\uparrow$} \\
        \midrule
        Random & 0.01\\
        \midrule
        Multi-modal & 0.70 \\
        + Contrastive & 0.83\\
        + Transformer  & 0.85 \\
        + LLM feedback  & 0.90\\
        \bottomrule
        \end{tabular}
        \vspace{1em}
        \captionof{table}{Ablation study on scenario recognition.}
        \label{tab:ab_scenario}
    \end{minipage}
    \vspace{-0.5em}
\end{figure}

Recalled in Sec. \ref{sec:scenario}, we employ a two-stage model to recognize the scenario by extracting the key-frame. In comparison, we consider audio scene recognition and ImageBind-IMU, which are audio-only and IMU-only methods
as the two baselines shown in Fig. \ref{fig:scenario}. We observe that our method outperforms the two baselines by 15\%, indicating that our long-time multi-modal fusion works well for scenario recognition. 

\para{Data length.}
Based on our earlier definitions, scenarios that require capturing long-term information should use longer data lengths. We present the performance of scenario recognition along with various data lengths in Fig. \ref{fig:scenario_length}. We observe that the ideal data length is set to 30 seconds since the larger input data can reduce the performance.

\para{Selection of LLM.}
Since our design is compatible with any LLM, we evaluated the performance using different LLMs from various sources. As shown in Fig. \ref{fig:selection_llm}, we observe that the performance is similar across them, indicating that common LLMs are sufficient for our design. For better deployment, we plan to explore the use of smaller LLMs in the future to reduce costs.

\para{Ablation study.}
We conducted an ablation study to evaluate the contribution of different components in our system, including contrastive learning, the transformer module, and the addition of LLM feedback. The results are presented in Tab. \ref{tab:ab_scenario}. For reference, we also include the performance of a random guess (based on the number of classes).

Our observations indicate that contrastive learning is the most effective component, providing the largest performance improvement (13\%). The LLM feedback, facilitated by edge-cloud collaboration, yields a further, albeit smaller, gain. Given that LLM feedback is less susceptible to cross-domain variations, we consider this slight improvement to be meaningful.

\subsection{Spatial Understanding for Scenario Recognition}
\begin{figure}[h]
    \centering
    \begin{minipage}{0.48\linewidth}
         \centering
    \includegraphics[width=1\linewidth]{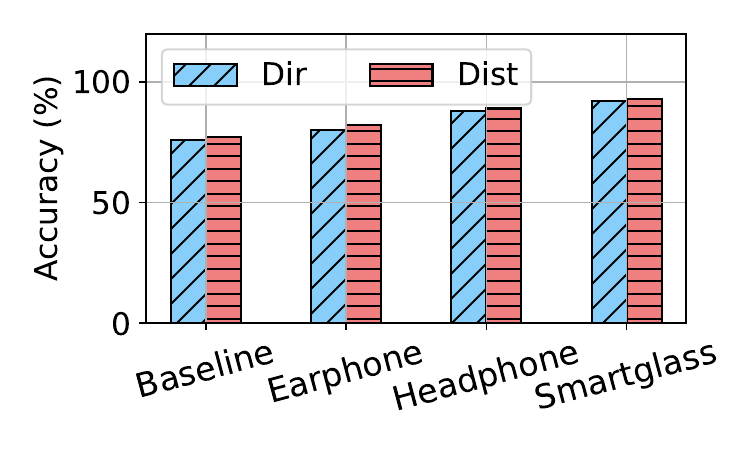}
            \vspace{-0.5em}
      \caption{Sound localization accuracy.}
        \label{fig:sound_localization}
    \end{minipage}
    \hfill
    \begin{minipage}{0.48\linewidth}
        \centering
    \includegraphics[width=1\linewidth]{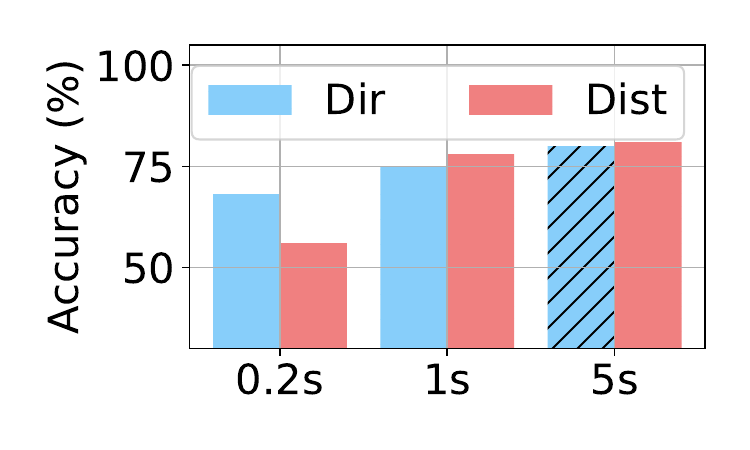}
          \vspace{-0.5em}
    \caption{Sound localization accuracy vs. data length.}
    \label{fig:sound_localization_length}
    \end{minipage}
    \vspace{-0.5em}
\end{figure}

\begin{figure}[h]
  \centering
  \begin{minipage}{0.4\linewidth}
    \centering
    \includegraphics[width=1\linewidth]{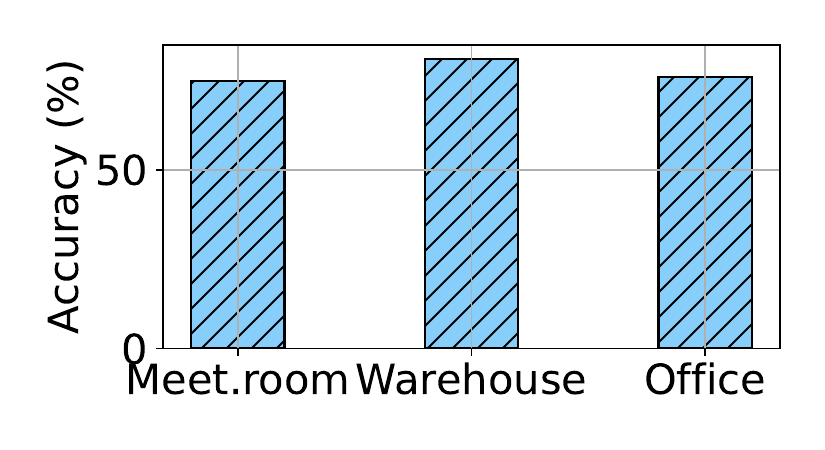}
        \vspace{-0.5em}
      \caption{Sound localization accuracy at different places.}
        \label{fig:sound_localization_place}
  \end{minipage}
  \hfill
  \begin{minipage}{0.58\linewidth}
    \centering
     \includegraphics[width=1\linewidth]{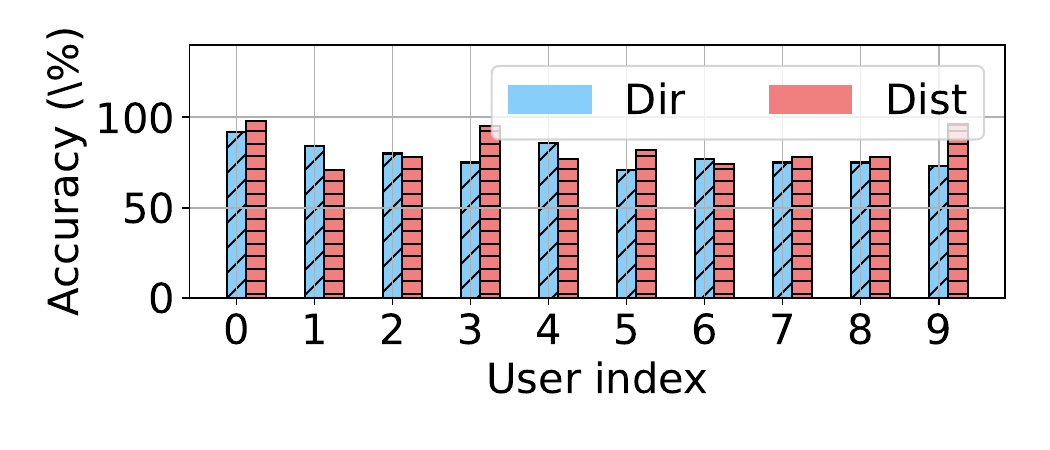}
         \vspace{-0.5em}
      \caption{Binaural sound localization accuracy for different users.}
        \label{fig:sound_localization_user}
  \end{minipage}
\vspace{-0.5em}
\end{figure}
In Fig. \ref{fig:sound_localization}, we compare the baseline (without motion compensation) with our proposed method across different devices. We report the accuracy for both direction (Dir) and distance (Dist) separately. The results show that our method improves accuracy by approximately 10\%when using IMU data for motion compensation.

Regarding performance across devices, we observe a consistent improvement from binaural earphones to 4-channel headphones and further to 4-channel smart glasses, which aligns with our expectations. In the following section, we will focus on evaluating data from earphones, which exhibit the lowest performance.

\para{Data length.}
Since the length of the audio significantly influences the performance of sound localization, we test the accuracy with different length of data in Fig. \ref{fig:sound_localization_length}. According to the result, we set the data length of sound localization to one second to balance the latency and accuracy.

\para{Different environments.}
We conducted further assessments of localization accuracy across various environments, such as distinct rooms. Room layout and object configurations induce varying room impulse responses, affecting recordings in conjunction with HRTFs. Our performance evaluation across different rooms is presented in Figure \ref{fig:sound_localization_place}, demonstrating that our model maintains consistent performance across diverse settings.

\para{Different users.}
Binaural sound localization depends on individual-specific HRTFs, rendering neural networks calibrated for one user
ineffective for others. We assess personalized models across a group of users and present the overall accuracy in Figure 21. Despite performance variations among users, consistent accuracy (>75\%) is maintained within the group

\subsection{User Study}


\new{
To evaluate the benefits of \systemname{} from a user perspective, we conducted a study where volunteers provided subjective ratings of the system's outputs. Recall that our goal is to provide users with a daily log that comprehensively summarizes and represents their activities through two complementary types of output: scenarios and activities. 
Our final solution integrates both outputs: the daily log displays two rows, with the first row showing scenario information and the second row presenting activity data. Since scenarios are recognized every 30 seconds, the output can contain errors or inconsistencies, even within the same scenario. To mitigate this, we represent each scenario using the normalized probability calculated from the most recent 10 minutes of data. 
For activities, due to the larger volume of data, we present the results using a bar chart with activity names labeled at the bottom.
Users were asked to rate the outputs based on three criteria: Correctness, Information-richness, and User-friendliness. To assess correctness, they were provided with ground-truth data in the same format. The average ratings (on a scale of 1-min to 5-max) for the three dimensions were 4.5, 4.0, and 4.5, respectively.
}

\subsection{Overhead}
While our system is not specifically designed for real-time responses, it is crucial to maintain low resource consumption.

\begin{table}[h]
\centering
\vspace{-0.5em}
\caption{Model runtime latency.}
\vspace{-0.5em}
\label{tab:latency}
\begin{tabular}{llcc}
\toprule
\textbf{Section} & \textbf{Interval} & \textbf{CPU} & \textbf{GPU} \\
\midrule
Scenario & 30 seconds & 0.55s & 9ms \\
Spatial & 2 seconds & 2.5ms & 2.1ms \\
HAR  & 10 seconds & 34ms & 2.7ms \\
\bottomrule
\end{tabular}
\vspace{-0.5em}
\end{table}

\para{Latency.}
\systemname{} comprises three core components: (1) scenario recognition (Sec. \ref{sec:scenario}), (2) sound localization (Sec. \ref{sec:localization}), and (3) HAR (Sec. \ref{sec:har}). As shown in Tab. \ref{tab:latency}, most components are lightweight, enabling real-time performance on both CPU and GPU platforms. Notably, scenario recognition on CPU requires 0.55 seconds but is executed once every 30 seconds, rendering its latency negligible. While \systemname{} does not mandate real-time user feedback, lower latency enhances user experience by minimizing wait times.

\para{Power consumption.}
To evaluate our power consumption, we locked the screen while continuously recording both audio and IMU data. We then checked the battery level before and after one hour of recording. Using the Pixel 7 as a reference, we observed approximately a 5\% drop in power after one hour of recording. Given that there is potential for further optimization, \systemname{} shows promise for effective long-term operation.

\para{Data storage.}
We record audio at a sample rate of 48 kHz, with a 16-bit depth and 2 channels, resulting in a compressed file size of about 170-210 kbps in MP3 format. This equates to around 70 MB of audio data per hour. For motion data, with an update rate set to 10 Hz, the size in CSV format is approximately 1 kB per second, leading to around 3 MB of data per hour.
As a result, we can process this data periodically, ensuring that storage requirements remain manageable and do not become overwhelming.

\section{Discussion}
\label{sec:discussion}
\para{Speech processing.}
While \systemname{} uses audio as a key input, it treats speech as a general sound class without specialized processing. Although automatic speech recognition (ASR) could extract speech content, its computational cost and reliance on cloud services pose privacy risks from potentially captured sensitive data.
Prior solutions like SAMOSA \cite{mollyn2022samosa} preserve privacy by degrading audio quality (e.g., reducing sample rates), but this sacrifices valuable acoustic details. Instead, we argue that for activity logging, full transcription is often unnecessary. A more efficient and private approach is on-device keyword spotting (KWS). An adaptive or open-vocabulary KWS system \cite{navon2024open} can detect relevant terms without comprehensive ASR. By leveraging an LLM to dynamically identify scenarioually important keywords, the system can provide effective support while ensuring user privacy.

\para{Beyond recognition.}
\systemname{} is primarily designed for activity and scenario recognition towards daily logging, as a classification problem.
The usability can be enhanced mainly in two ways: 1) zero-shot or free-form recognition, and 2) recognition with feedback.
For zero-shot recognition, extending language models to process multi-modal sensor data (e.g., audio, IMU) presents difficulties due to data sparsity. The core challenge lies in determining appropriate recognition granularity—more detailed than basic classification but within the capability, and constructing suitable datasets to support this capability.
For feedback integration, we can evolve from passive logging to proactive assistance. This enables advanced applications such as social support and context-aware agent services, providing users with intelligent, scenario-relevant guidance and historical interaction summaries.

\para{Large-scale Dataset.}
Our self-collected dataset lacks the diversity of Ego4D \cite{grauman2022ego4d}, but collecting a similarly large dataset with spatial audio and motion, as attempted by \cite{min2025supervising} using multichannel YouTube audio, is promising. Annotating such datasets, especially for spatial audio sources, is challenging and time-consuming, often requiring a lab setting, such as attaching a tracking marker or sensor to all the potential sources. Inspired by \cite{min2025supervising}'s use of noisy vision-based ground-truth for efficient annotation, we envision collecting a new egocentric audio dataset using commercial and advanced audio devices for accurate sound source estimation.

\section{Conclusion}
In conclusion, \systemname{} advances daily activity logging through a novel audio-IMU fusion technique that leverages scenario and spatial awareness. Our scenario-aware, multi-modal framework significantly enhances recognition granularity, outperforming existing solutions. This performance is further boosted by an efficient edge-cloud collaboration, where a cloud-based LLM refines scenario recognition. Evaluated on both public and self-collected datasets using ubiquitous wearables, \systemname{} demonstrates its practicality as a robust solution for personalized healthcare and long-term monitoring.

\newpage
\bibliographystyle{ACM-Reference-Format}
\bibliography{sample-base}


\begin{thebibliography}{56}


\ifx \showCODEN    \undefined \def \showCODEN     #1{\unskip}     \fi
\ifx \showISBNx    \undefined \def \showISBNx     #1{\unskip}     \fi
\ifx \showISBNxiii \undefined \def \showISBNxiii  #1{\unskip}     \fi
\ifx \showISSN     \undefined \def \showISSN      #1{\unskip}     \fi
\ifx \showLCCN     \undefined \def \showLCCN      #1{\unskip}     \fi
\ifx \shownote     \undefined \def \shownote      #1{#1}          \fi
\ifx \showarticletitle \undefined \def \showarticletitle #1{#1}   \fi
\ifx \showURL      \undefined \def \showURL       {\relax}        \fi
\providecommand\bibfield[2]{#2}
\providecommand\bibinfo[2]{#2}
\providecommand\natexlab[1]{#1}
\providecommand\showeprint[2][]{arXiv:#2}

\bibitem[sou(2015)]%
        {soundprofessionals}
 \bibinfo{year}{2015}\natexlab{}.
\newblock
\urldef\tempurl%
\url{https://soundprofessionals.com/}
\showURL{%
\tempurl}


\bibitem[Mu6(2020)]%
        {Mu6_Home_Page_2020}
 \bibinfo{year}{2020}\natexlab{}.
\newblock
\urldef\tempurl%
\url{https://www.mu6.live/}
\showURL{%
\tempurl}


\bibitem[Adavanne et~al\mbox{.}(2018)]%
        {adavanne2018sound}
\bibfield{author}{\bibinfo{person}{Sharath Adavanne}, \bibinfo{person}{Archontis Politis}, \bibinfo{person}{Joonas Nikunen}, {and} \bibinfo{person}{Tuomas Virtanen}.} \bibinfo{year}{2018}\natexlab{}.
\newblock \showarticletitle{Sound event localization and detection of overlapping sources using convolutional recurrent neural networks}.
\newblock \bibinfo{journal}{\emph{IEEE Journal of Selected Topics in Signal Processing}} \bibinfo{volume}{13}, \bibinfo{number}{1} (\bibinfo{year}{2018}), \bibinfo{pages}{34--48}.
\newblock


\bibitem[Alam and Roy(2014)]%
        {alam2014gesmart}
\bibfield{author}{\bibinfo{person}{Mohammad Arif~Ul Alam} {and} \bibinfo{person}{Nirmalya Roy}.} \bibinfo{year}{2014}\natexlab{}.
\newblock \showarticletitle{GeSmart: A gestural activity recognition model for predicting behavioral health}. In \bibinfo{booktitle}{\emph{2014 International Conference on Smart Computing}}. IEEE, \bibinfo{pages}{193--200}.
\newblock


\bibitem[Bhattacharya et~al\mbox{.}(2022)]%
        {bhattacharya2022leveraging}
\bibfield{author}{\bibinfo{person}{Sarnab Bhattacharya}, \bibinfo{person}{Rebecca Adaimi}, {and} \bibinfo{person}{Edison Thomaz}.} \bibinfo{year}{2022}\natexlab{}.
\newblock \showarticletitle{Leveraging sound and wrist motion to detect activities of daily living with commodity smartwatches}.
\newblock \bibinfo{journal}{\emph{Proceedings of the ACM on Interactive, Mobile, Wearable and Ubiquitous Technologies}} \bibinfo{volume}{6}, \bibinfo{number}{2} (\bibinfo{year}{2022}), \bibinfo{pages}{1--28}.
\newblock


\bibitem[Chan et~al\mbox{.}(2024)]%
        {chan2024capture}
\bibfield{author}{\bibinfo{person}{Shing Chan}, \bibinfo{person}{Yuan Hang}, \bibinfo{person}{Catherine Tong}, \bibinfo{person}{Aidan Acquah}, \bibinfo{person}{Abram Schonfeldt}, \bibinfo{person}{Jonathan Gershuny}, {and} \bibinfo{person}{Aiden Doherty}.} \bibinfo{year}{2024}\natexlab{}.
\newblock \showarticletitle{CAPTURE-24: A large dataset of wrist-worn activity tracker data collected in the wild for human activity recognition}.
\newblock \bibinfo{journal}{\emph{Scientific Data}} \bibinfo{volume}{11}, \bibinfo{number}{1} (\bibinfo{year}{2024}), \bibinfo{pages}{1135}.
\newblock


\bibitem[Chen et~al\mbox{.}(2024)]%
        {chen2024hearable}
\bibfield{author}{\bibinfo{person}{Tuochao Chen}, \bibinfo{person}{Malek Itani}, \bibinfo{person}{Sefik~Emre Eskimez}, \bibinfo{person}{Takuya Yoshioka}, {and} \bibinfo{person}{Shyamnath Gollakota}.} \bibinfo{year}{2024}\natexlab{}.
\newblock \showarticletitle{Hearable devices with sound bubbles}.
\newblock \bibinfo{journal}{\emph{Nature Electronics}} (\bibinfo{year}{2024}), \bibinfo{pages}{1--12}.
\newblock


\bibitem[Chen et~al\mbox{.}(2020)]%
        {chen2020simple}
\bibfield{author}{\bibinfo{person}{Ting Chen}, \bibinfo{person}{Simon Kornblith}, \bibinfo{person}{Mohammad Norouzi}, {and} \bibinfo{person}{Geoffrey Hinton}.} \bibinfo{year}{2020}\natexlab{}.
\newblock \showarticletitle{A simple framework for contrastive learning of visual representations}. In \bibinfo{booktitle}{\emph{International conference on machine learning}}. PmLR, \bibinfo{pages}{1597--1607}.
\newblock


\bibitem[Cristina et~al\mbox{.}(2024)]%
        {cristina2024audio}
\bibfield{author}{\bibinfo{person}{Stefania Cristina}, \bibinfo{person}{Vladimir Despotovic}, \bibinfo{person}{Rodrigo P{\'e}rez-Rodr{\'\i}guez}, {and} \bibinfo{person}{Slavisa Aleksic}.} \bibinfo{year}{2024}\natexlab{}.
\newblock \showarticletitle{Audio-and video-based human activity recognition systems in healthcare}.
\newblock \bibinfo{journal}{\emph{IEEE Access}}  \bibinfo{volume}{12} (\bibinfo{year}{2024}), \bibinfo{pages}{8230--8245}.
\newblock


\bibitem[Dong et~al\mbox{.}(2024)]%
        {dong2024rehearsse}
\bibfield{author}{\bibinfo{person}{Xuefu Dong}, \bibinfo{person}{Yifei Chen}, \bibinfo{person}{Yuuki Nishiyama}, \bibinfo{person}{Kaoru Sezaki}, \bibinfo{person}{Yuntao Wang}, \bibinfo{person}{Ken Christofferson}, {and} \bibinfo{person}{Alex Mariakakis}.} \bibinfo{year}{2024}\natexlab{}.
\newblock \showarticletitle{ReHEarSSE: Recognizing Hidden-in-the-Ear Silently Spelled Expressions}. In \bibinfo{booktitle}{\emph{Proceedings of the CHI Conference on Human Factors in Computing Systems}}. \bibinfo{pages}{1--16}.
\newblock


\bibitem[Gabeur et~al\mbox{.}(2020)]%
        {gabeur2020multi}
\bibfield{author}{\bibinfo{person}{Valentin Gabeur}, \bibinfo{person}{Chen Sun}, \bibinfo{person}{Karteek Alahari}, {and} \bibinfo{person}{Cordelia Schmid}.} \bibinfo{year}{2020}\natexlab{}.
\newblock \showarticletitle{Multi-modal transformer for video retrieval}. In \bibinfo{booktitle}{\emph{Computer Vision--ECCV 2020: 16th European Conference, Glasgow, UK, August 23--28, 2020, Proceedings, Part IV 16}}. Springer, \bibinfo{pages}{214--229}.
\newblock


\bibitem[Gedam and Paul(2021)]%
        {gedam2021review}
\bibfield{author}{\bibinfo{person}{Shruti Gedam} {and} \bibinfo{person}{Sanchita Paul}.} \bibinfo{year}{2021}\natexlab{}.
\newblock \showarticletitle{A review on mental stress detection using wearable sensors and machine learning techniques}.
\newblock \bibinfo{journal}{\emph{IEEE Access}}  \bibinfo{volume}{9} (\bibinfo{year}{2021}), \bibinfo{pages}{84045--84066}.
\newblock


\bibitem[Girdhar et~al\mbox{.}(2023)]%
        {girdhar2023imagebind}
\bibfield{author}{\bibinfo{person}{Rohit Girdhar}, \bibinfo{person}{Alaaeldin El-Nouby}, \bibinfo{person}{Zhuang Liu}, \bibinfo{person}{Mannat Singh}, \bibinfo{person}{Kalyan~Vasudev Alwala}, \bibinfo{person}{Armand Joulin}, {and} \bibinfo{person}{Ishan Misra}.} \bibinfo{year}{2023}\natexlab{}.
\newblock \showarticletitle{Imagebind: One embedding space to bind them all}. In \bibinfo{booktitle}{\emph{Proceedings of the IEEE/CVF Conference on Computer Vision and Pattern Recognition}}. \bibinfo{pages}{15180--15190}.
\newblock


\bibitem[Gong et~al\mbox{.}(2023)]%
        {gong2023mmg}
\bibfield{author}{\bibinfo{person}{Xinyu Gong}, \bibinfo{person}{Sreyas Mohan}, \bibinfo{person}{Naina Dhingra}, \bibinfo{person}{Jean-Charles Bazin}, \bibinfo{person}{Yilei Li}, \bibinfo{person}{Zhangyang Wang}, {and} \bibinfo{person}{Rakesh Ranjan}.} \bibinfo{year}{2023}\natexlab{}.
\newblock \showarticletitle{MMG-ego4D: multimodal generalization in egocentric action recognition}. In \bibinfo{booktitle}{\emph{Proceedings of the IEEE/CVF Conference on Computer Vision and Pattern Recognition}}. \bibinfo{pages}{6481--6491}.
\newblock


\bibitem[Grauman et~al\mbox{.}(2022)]%
        {grauman2022ego4d}
\bibfield{author}{\bibinfo{person}{Kristen Grauman}, \bibinfo{person}{Andrew Westbury}, \bibinfo{person}{Eugene Byrne}, \bibinfo{person}{Zachary Chavis}, \bibinfo{person}{Antonino Furnari}, \bibinfo{person}{Rohit Girdhar}, \bibinfo{person}{Jackson Hamburger}, \bibinfo{person}{Hao Jiang}, \bibinfo{person}{Miao Liu}, \bibinfo{person}{Xingyu Liu}, {et~al\mbox{.}}} \bibinfo{year}{2022}\natexlab{}.
\newblock \showarticletitle{Ego4d: Around the world in 3,000 hours of egocentric video}. In \bibinfo{booktitle}{\emph{Proceedings of the IEEE/CVF Conference on Computer Vision and Pattern Recognition}}. \bibinfo{pages}{18995--19012}.
\newblock


\bibitem[Han et~al\mbox{.}(2023b)]%
        {han2023onellm}
\bibfield{author}{\bibinfo{person}{Jiaming Han}, \bibinfo{person}{Kaixiong Gong}, \bibinfo{person}{Yiyuan Zhang}, \bibinfo{person}{Jiaqi Wang}, \bibinfo{person}{Kaipeng Zhang}, \bibinfo{person}{Dahua Lin}, \bibinfo{person}{Yu Qiao}, \bibinfo{person}{Peng Gao}, {and} \bibinfo{person}{Xiangyu Yue}.} \bibinfo{year}{2023}\natexlab{b}.
\newblock \showarticletitle{Onellm: One framework to align all modalities with language}.
\newblock \bibinfo{journal}{\emph{arXiv preprint arXiv:2312.03700}} (\bibinfo{year}{2023}).
\newblock


\bibitem[Han et~al\mbox{.}(2023d)]%
        {han2023imagebind}
\bibfield{author}{\bibinfo{person}{Jiaming Han}, \bibinfo{person}{Renrui Zhang}, \bibinfo{person}{Wenqi Shao}, \bibinfo{person}{Peng Gao}, \bibinfo{person}{Peng Xu}, \bibinfo{person}{Han Xiao}, \bibinfo{person}{Kaipeng Zhang}, \bibinfo{person}{Chris Liu}, \bibinfo{person}{Song Wen}, \bibinfo{person}{Ziyu Guo}, {et~al\mbox{.}}} \bibinfo{year}{2023}\natexlab{d}.
\newblock \showarticletitle{Imagebind-llm: Multi-modality instruction tuning}.
\newblock \bibinfo{journal}{\emph{arXiv preprint arXiv:2309.03905}} (\bibinfo{year}{2023}).
\newblock


\bibitem[Han et~al\mbox{.}(2023a)]%
        {han2023headsense}
\bibfield{author}{\bibinfo{person}{Zengyi Han}, \bibinfo{person}{Xuefu Dong}, \bibinfo{person}{Yuuki Nishiyama}, {and} \bibinfo{person}{Kaoru Sezaki}.} \bibinfo{year}{2023}\natexlab{a}.
\newblock \showarticletitle{HeadSense: Visual Search Monitoring and Distracted Behavior Detection for Bicycle Riders}. In \bibinfo{booktitle}{\emph{2023 IEEE 24th International Symposium on a World of Wireless, Mobile and Multimedia Networks (WoWMoM)}}. IEEE, \bibinfo{pages}{281--289}.
\newblock


\bibitem[Han et~al\mbox{.}(2025)]%
        {han2025headmon+}
\bibfield{author}{\bibinfo{person}{Zengyi Han}, \bibinfo{person}{En Wang}, \bibinfo{person}{Mohan Yu}, \bibinfo{person}{Jie Wang}, \bibinfo{person}{Yuuki Nishiyama}, {and} \bibinfo{person}{Kaoru Sezaki}.} \bibinfo{year}{2025}\natexlab{}.
\newblock \showarticletitle{HeadMon+: Domain Adaptive Head Dynamic-based Riding Maneuver Prediction}.
\newblock \bibinfo{journal}{\emph{IEEE Transactions on Mobile Computing}} (\bibinfo{year}{2025}).
\newblock


\bibitem[Han et~al\mbox{.}(2023c)]%
        {han2023headmon}
\bibfield{author}{\bibinfo{person}{Zengyi Han}, \bibinfo{person}{Liqiang Xu}, \bibinfo{person}{Xuefu Dong}, \bibinfo{person}{Yuuki Nishiyama}, {and} \bibinfo{person}{Kaoru Sezaki}.} \bibinfo{year}{2023}\natexlab{c}.
\newblock \showarticletitle{Headmon: Head dynamics enabled riding maneuver prediction}. In \bibinfo{booktitle}{\emph{2023 IEEE International Conference on Pervasive Computing and Communications (PerCom)}}. IEEE, \bibinfo{pages}{22--31}.
\newblock


\bibitem[He et~al\mbox{.}(2023)]%
        {he2023towards}
\bibfield{author}{\bibinfo{person}{Lixing He}, \bibinfo{person}{Haozheng Hou}, \bibinfo{person}{Shuyao Shi}, \bibinfo{person}{Xian Shuai}, {and} \bibinfo{person}{Zhenyu Yan}.} \bibinfo{year}{2023}\natexlab{}.
\newblock \showarticletitle{Towards Bone-Conducted Vibration Speech Enhancement on Head-Mounted Wearables}. In \bibinfo{booktitle}{\emph{Proceedings of the 21st Annual International Conference on Mobile Systems, Applications and Services}}. \bibinfo{pages}{14--27}.
\newblock


\bibitem[Huang et~al\mbox{.}(2022)]%
        {huang2022modality}
\bibfield{author}{\bibinfo{person}{Yu Huang}, \bibinfo{person}{Junyang Lin}, \bibinfo{person}{Chang Zhou}, \bibinfo{person}{Hongxia Yang}, {and} \bibinfo{person}{Longbo Huang}.} \bibinfo{year}{2022}\natexlab{}.
\newblock \showarticletitle{Modality competition: What makes joint training of multi-modal network fail in deep learning?(provably)}. In \bibinfo{booktitle}{\emph{International Conference on Machine Learning}}. PMLR, \bibinfo{pages}{9226--9259}.
\newblock


\bibitem[Ji et~al\mbox{.}(2024)]%
        {ji2024hargpt}
\bibfield{author}{\bibinfo{person}{Sijie Ji}, \bibinfo{person}{Xinzhe Zheng}, {and} \bibinfo{person}{Chenshu Wu}.} \bibinfo{year}{2024}\natexlab{}.
\newblock \showarticletitle{HARGPT: Are LLMs Zero-Shot Human Activity Recognizers?}
\newblock \bibinfo{journal}{\emph{arXiv preprint arXiv:2403.02727}} (\bibinfo{year}{2024}).
\newblock


\bibitem[Krause et~al\mbox{.}(2021)]%
        {krause2021joint}
\bibfield{author}{\bibinfo{person}{Daniel~Aleksander Krause}, \bibinfo{person}{Archontis Politis}, {and} \bibinfo{person}{Annamaria Mesaros}.} \bibinfo{year}{2021}\natexlab{}.
\newblock \showarticletitle{Joint direction and proximity classification of overlapping sound events from binaural audio}. In \bibinfo{booktitle}{\emph{2021 IEEE Workshop on Applications of Signal Processing to Audio and Acoustics (WASPAA)}}. IEEE, \bibinfo{pages}{331--335}.
\newblock


\bibitem[Leng et~al\mbox{.}(2024)]%
        {leng2024imugpt}
\bibfield{author}{\bibinfo{person}{Zikang Leng}, \bibinfo{person}{Amitrajit Bhattacharjee}, \bibinfo{person}{Hrudhai Rajasekhar}, \bibinfo{person}{Lizhe Zhang}, \bibinfo{person}{Elizabeth Bruda}, \bibinfo{person}{Hyeokhyen Kwon}, {and} \bibinfo{person}{Thomas Pl{\"o}tz}.} \bibinfo{year}{2024}\natexlab{}.
\newblock \showarticletitle{Imugpt 2.0: Language-based cross modality transfer for sensor-based human activity recognition}.
\newblock \bibinfo{journal}{\emph{Proceedings of the ACM on Interactive, Mobile, Wearable and Ubiquitous Technologies}} \bibinfo{volume}{8}, \bibinfo{number}{3} (\bibinfo{year}{2024}), \bibinfo{pages}{1--32}.
\newblock


\bibitem[Lyu et~al\mbox{.}(2024)]%
        {lyu2024earda}
\bibfield{author}{\bibinfo{person}{Shengzhe Lyu}, \bibinfo{person}{Yongliang Chen}, \bibinfo{person}{Di Duan}, \bibinfo{person}{Renqi Jia}, {and} \bibinfo{person}{Weitao Xu}.} \bibinfo{year}{2024}\natexlab{}.
\newblock \showarticletitle{Earda: Towards accurate and data-efficient earable activity sensing}. In \bibinfo{booktitle}{\emph{2024 IEEE Coupling of Sensing \& Computing in AIoT Systems (CSCAIoT)}}. IEEE, \bibinfo{pages}{1--7}.
\newblock


\bibitem[MacQueen(1967)]%
        {macqueen1967some}
\bibfield{author}{\bibinfo{person}{J MacQueen}.} \bibinfo{year}{1967}\natexlab{}.
\newblock \showarticletitle{Some methods for classification and analysis of multivariate observations}. In \bibinfo{booktitle}{\emph{Proceedings of 5-th Berkeley Symposium on Mathematical Statistics and Probability/University of California Press}}.
\newblock


\bibitem[Mahmud et~al\mbox{.}(2023)]%
        {mahmud2023posesonic}
\bibfield{author}{\bibinfo{person}{Saif Mahmud}, \bibinfo{person}{Ke Li}, \bibinfo{person}{Guilin Hu}, \bibinfo{person}{Hao Chen}, \bibinfo{person}{Richard Jin}, \bibinfo{person}{Ruidong Zhang}, \bibinfo{person}{Fran{\c{c}}ois Guimbreti{\`e}re}, {and} \bibinfo{person}{Cheng Zhang}.} \bibinfo{year}{2023}\natexlab{}.
\newblock \showarticletitle{Posesonic: 3d upper body pose estimation through egocentric acoustic sensing on smartglasses}.
\newblock \bibinfo{journal}{\emph{Proceedings of the ACM on Interactive, Mobile, Wearable and Ubiquitous Technologies}} \bibinfo{volume}{7}, \bibinfo{number}{3} (\bibinfo{year}{2023}), \bibinfo{pages}{1--28}.
\newblock


\bibitem[Mart{\'\i}n-Morat{\'o} et~al\mbox{.}(2021)]%
        {martin2021low}
\bibfield{author}{\bibinfo{person}{Irene Mart{\'\i}n-Morat{\'o}}, \bibinfo{person}{Toni Heittola}, \bibinfo{person}{Annamaria Mesaros}, {and} \bibinfo{person}{Tuomas Virtanen}.} \bibinfo{year}{2021}\natexlab{}.
\newblock \showarticletitle{Low-complexity acoustic scene classification for multi-device audio: Analysis of DCASE 2021 challenge systems}.
\newblock \bibinfo{journal}{\emph{arXiv preprint arXiv:2105.13734}} (\bibinfo{year}{2021}).
\newblock


\bibitem[Min et~al\mbox{.}(2025)]%
        {min2025supervising}
\bibfield{author}{\bibinfo{person}{Anna Min}, \bibinfo{person}{Ziyang Chen}, \bibinfo{person}{Hang Zhao}, {and} \bibinfo{person}{Andrew Owens}.} \bibinfo{year}{2025}\natexlab{}.
\newblock \showarticletitle{Supervising Sound Localization by In-the-wild Egomotion}. In \bibinfo{booktitle}{\emph{Proceedings of the Computer Vision and Pattern Recognition Conference}}. \bibinfo{pages}{23936--23946}.
\newblock


\bibitem[Min et~al\mbox{.}(2018)]%
        {min2018exploring}
\bibfield{author}{\bibinfo{person}{Chulhong Min}, \bibinfo{person}{Akhil Mathur}, {and} \bibinfo{person}{Fahim Kawsar}.} \bibinfo{year}{2018}\natexlab{}.
\newblock \showarticletitle{Exploring audio and kinetic sensing on earable devices}. In \bibinfo{booktitle}{\emph{Proceedings of the 4th ACM Workshop on Wearable Systems and Applications}}. \bibinfo{pages}{5--10}.
\newblock


\bibitem[Mollyn et~al\mbox{.}(2022)]%
        {mollyn2022samosa}
\bibfield{author}{\bibinfo{person}{Vimal Mollyn}, \bibinfo{person}{Karan Ahuja}, \bibinfo{person}{Dhruv Verma}, \bibinfo{person}{Chris Harrison}, {and} \bibinfo{person}{Mayank Goel}.} \bibinfo{year}{2022}\natexlab{}.
\newblock \showarticletitle{SAMoSA: Sensing activities with motion and subsampled audio}.
\newblock \bibinfo{journal}{\emph{Proceedings of the ACM on Interactive, Mobile, Wearable and Ubiquitous Technologies}} \bibinfo{volume}{6}, \bibinfo{number}{3} (\bibinfo{year}{2022}), \bibinfo{pages}{1--19}.
\newblock


\bibitem[Mollyn et~al\mbox{.}(2023)]%
        {mollyn2023imuposer}
\bibfield{author}{\bibinfo{person}{Vimal Mollyn}, \bibinfo{person}{Riku Arakawa}, \bibinfo{person}{Mayank Goel}, \bibinfo{person}{Chris Harrison}, {and} \bibinfo{person}{Karan Ahuja}.} \bibinfo{year}{2023}\natexlab{}.
\newblock \showarticletitle{Imuposer: Full-body pose estimation using imus in phones, watches, and earbuds}. In \bibinfo{booktitle}{\emph{Proceedings of the 2023 CHI Conference on Human Factors in Computing Systems}}. \bibinfo{pages}{1--12}.
\newblock


\bibitem[Moon et~al\mbox{.}(2023)]%
        {moon2023imu2clip}
\bibfield{author}{\bibinfo{person}{Seungwhan Moon}, \bibinfo{person}{Andrea Madotto}, \bibinfo{person}{Zhaojiang Lin}, \bibinfo{person}{Aparajita Saraf}, \bibinfo{person}{Amy Bearman}, {and} \bibinfo{person}{Babak Damavandi}.} \bibinfo{year}{2023}\natexlab{}.
\newblock \showarticletitle{IMU2CLIP: Language-grounded Motion Sensor Translation with Multimodal Contrastive Learning}. In \bibinfo{booktitle}{\emph{Findings of the Association for Computational Linguistics: EMNLP 2023}}. \bibinfo{pages}{13246--13253}.
\newblock


\bibitem[Navon et~al\mbox{.}(2024)]%
        {navon2024open}
\bibfield{author}{\bibinfo{person}{Aviv Navon}, \bibinfo{person}{Aviv Shamsian}, \bibinfo{person}{Neta Glazer}, \bibinfo{person}{Gill Hetz}, {and} \bibinfo{person}{Joseph Keshet}.} \bibinfo{year}{2024}\natexlab{}.
\newblock \showarticletitle{Open-vocabulary keyword-spotting with adaptive instance normalization}. In \bibinfo{booktitle}{\emph{ICASSP 2024-2024 IEEE International Conference on Acoustics, Speech and Signal Processing (ICASSP)}}. IEEE, \bibinfo{pages}{11656--11660}.
\newblock


\bibitem[Ouyang and Srivastava(2024)]%
        {ouyang2024llmsense}
\bibfield{author}{\bibinfo{person}{Xiaomin Ouyang} {and} \bibinfo{person}{Mani Srivastava}.} \bibinfo{year}{2024}\natexlab{}.
\newblock \showarticletitle{LLMSense: Harnessing LLMs for High-level Reasoning Over Spatiotemporal Sensor Traces}.
\newblock \bibinfo{journal}{\emph{arXiv preprint arXiv:2403.19857}} (\bibinfo{year}{2024}).
\newblock


\bibitem[Reimers(2019)]%
        {reimers2019sentence}
\bibfield{author}{\bibinfo{person}{N Reimers}.} \bibinfo{year}{2019}\natexlab{}.
\newblock \showarticletitle{Sentence-BERT: Sentence Embeddings using Siamese BERT-Networks}.
\newblock \bibinfo{journal}{\emph{arXiv preprint arXiv:1908.10084}} (\bibinfo{year}{2019}).
\newblock


\bibitem[Reyes-Ortiz et~al\mbox{.}(2015)]%
        {smartphonebased_recognition_of_human_activities_and_postural_transitions_341}
\bibfield{author}{\bibinfo{person}{Jorge Reyes-Ortiz}, \bibinfo{person}{Davide Anguita}, \bibinfo{person}{Luca Oneto}, {and} \bibinfo{person}{Xavier Parra}.} \bibinfo{year}{2015}\natexlab{}.
\newblock \bibinfo{title}{{Smartphone-Based Recognition of Human Activities and Postural Transitions}}.
\newblock \bibinfo{howpublished}{UCI Machine Learning Repository}.
\newblock
\newblock
\shownote{{DOI}: https://doi.org/10.24432/C54G7M}.


\bibitem[Schmid et~al\mbox{.}(2023)]%
        {schmid2023efficient}
\bibfield{author}{\bibinfo{person}{Florian Schmid}, \bibinfo{person}{Khaled Koutini}, {and} \bibinfo{person}{Gerhard Widmer}.} \bibinfo{year}{2023}\natexlab{}.
\newblock \showarticletitle{Efficient large-scale audio tagging via transformer-to-cnn knowledge distillation}. In \bibinfo{booktitle}{\emph{ICASSP 2023-2023 IEEE International Conference on Acoustics, Speech and Signal Processing (ICASSP)}}. IEEE, \bibinfo{pages}{1--5}.
\newblock


\bibitem[Schmidt(1986)]%
        {schmidt1986multiple}
\bibfield{author}{\bibinfo{person}{Ralph Schmidt}.} \bibinfo{year}{1986}\natexlab{}.
\newblock \showarticletitle{Multiple emitter location and signal parameter estimation}.
\newblock \bibinfo{journal}{\emph{IEEE transactions on antennas and propagation}} \bibinfo{volume}{34}, \bibinfo{number}{3} (\bibinfo{year}{1986}), \bibinfo{pages}{276--280}.
\newblock


\bibitem[Seo et~al\mbox{.}(2022)]%
        {seo2022ross}
\bibfield{author}{\bibinfo{person}{Hyungjoo Seo}, \bibinfo{person}{Sahil~Bhandary Karnoor}, {and} \bibinfo{person}{Romit~Roy Choudhury}.} \bibinfo{year}{2022}\natexlab{}.
\newblock \showarticletitle{RoSS: Utilizing Robotic Rotation for Audio Source Separation}.
\newblock \bibinfo{journal}{\emph{arXiv preprint arXiv:2203.10072}} (\bibinfo{year}{2022}).
\newblock


\bibitem[Str{\"o}mb{\"a}ck et~al\mbox{.}(2020)]%
        {stromback2020mm}
\bibfield{author}{\bibinfo{person}{David Str{\"o}mb{\"a}ck}, \bibinfo{person}{Sangxia Huang}, {and} \bibinfo{person}{Valentin Radu}.} \bibinfo{year}{2020}\natexlab{}.
\newblock \showarticletitle{Mm-fit: Multimodal deep learning for automatic exercise logging across sensing devices}.
\newblock \bibinfo{journal}{\emph{Proceedings of the ACM on Interactive, Mobile, Wearable and Ubiquitous Technologies}} \bibinfo{volume}{4}, \bibinfo{number}{4} (\bibinfo{year}{2020}), \bibinfo{pages}{1--22}.
\newblock


\bibitem[Sun et~al\mbox{.}(2024)]%
        {sun2024multimodal}
\bibfield{author}{\bibinfo{person}{Ke Sun}, \bibinfo{person}{Chunyu Xia}, \bibinfo{person}{Xinyu Zhang}, \bibinfo{person}{Hao Chen}, {and} \bibinfo{person}{Charlie~Jianzhong Zhang}.} \bibinfo{year}{2024}\natexlab{}.
\newblock \showarticletitle{Multimodal daily-life logging in free-living environment using non-visual egocentric sensors on a smartphone}.
\newblock \bibinfo{journal}{\emph{Proceedings of the ACM on Interactive, Mobile, Wearable and Ubiquitous Technologies}} \bibinfo{volume}{8}, \bibinfo{number}{1} (\bibinfo{year}{2024}), \bibinfo{pages}{1--32}.
\newblock


\bibitem[Tong et~al\mbox{.}(2020)]%
        {tong2020accelerometers}
\bibfield{author}{\bibinfo{person}{Catherine Tong}, \bibinfo{person}{Shyam~A Tailor}, {and} \bibinfo{person}{Nicholas~D Lane}.} \bibinfo{year}{2020}\natexlab{}.
\newblock \showarticletitle{Are accelerometers for activity recognition a dead-end?}. In \bibinfo{booktitle}{\emph{Proceedings of the 21st international workshop on mobile computing systems and applications}}. \bibinfo{pages}{39--44}.
\newblock


\bibitem[Trowitzsch et~al\mbox{.}(2019)]%
        {trowitzsch2019nigens}
\bibfield{author}{\bibinfo{person}{Ivo Trowitzsch}, \bibinfo{person}{Jalil Taghia}, \bibinfo{person}{Youssef Kashef}, {and} \bibinfo{person}{Klaus Obermayer}.} \bibinfo{year}{2019}\natexlab{}.
\newblock \showarticletitle{The NIGENS general sound events database}.
\newblock \bibinfo{journal}{\emph{arXiv preprint arXiv:1902.08314}} (\bibinfo{year}{2019}).
\newblock


\bibitem[Xu et~al\mbox{.}(2024a)]%
        {xu2024penetrative}
\bibfield{author}{\bibinfo{person}{Huatao Xu}, \bibinfo{person}{Liying Han}, \bibinfo{person}{Qirui Yang}, \bibinfo{person}{Mo Li}, {and} \bibinfo{person}{Mani Srivastava}.} \bibinfo{year}{2024}\natexlab{a}.
\newblock \showarticletitle{Penetrative ai: Making llms comprehend the physical world}. In \bibinfo{booktitle}{\emph{Proceedings of the 25th International Workshop on Mobile Computing Systems and Applications}}. \bibinfo{pages}{1--7}.
\newblock


\bibitem[Xu et~al\mbox{.}(2024b)]%
        {xu2024autolife}
\bibfield{author}{\bibinfo{person}{Huatao Xu}, \bibinfo{person}{Panrong Tong}, \bibinfo{person}{Mo Li}, {and} \bibinfo{person}{Mani Srivastava}.} \bibinfo{year}{2024}\natexlab{b}.
\newblock \showarticletitle{AutoLife: Automatic Life Journaling with Smartphones and LLMs}.
\newblock \bibinfo{journal}{\emph{arXiv preprint arXiv:2412.15714}} (\bibinfo{year}{2024}).
\newblock


\bibitem[Xu et~al\mbox{.}(2021)]%
        {xu2021limu}
\bibfield{author}{\bibinfo{person}{Huatao Xu}, \bibinfo{person}{Pengfei Zhou}, \bibinfo{person}{Rui Tan}, \bibinfo{person}{Mo Li}, {and} \bibinfo{person}{Guobin Shen}.} \bibinfo{year}{2021}\natexlab{}.
\newblock \showarticletitle{Limu-bert: Unleashing the potential of unlabeled data for imu sensing applications}. In \bibinfo{booktitle}{\emph{Proceedings of the 19th ACM Conference on Embedded Networked Sensor Systems}}. \bibinfo{pages}{220--233}.
\newblock


\bibitem[Yang et~al\mbox{.}(2024b)]%
        {yang2024drhouse}
\bibfield{author}{\bibinfo{person}{Bufang Yang}, \bibinfo{person}{Siyang Jiang}, \bibinfo{person}{Lilin Xu}, \bibinfo{person}{Kaiwei Liu}, \bibinfo{person}{Hai Li}, \bibinfo{person}{Guoliang Xing}, \bibinfo{person}{Hongkai Chen}, \bibinfo{person}{Xiaofan Jiang}, {and} \bibinfo{person}{Zhenyu Yan}.} \bibinfo{year}{2024}\natexlab{b}.
\newblock \showarticletitle{Drhouse: An llm-empowered diagnostic reasoning system through harnessing outcomes from sensor data and expert knowledge}.
\newblock \bibinfo{journal}{\emph{Proceedings of the ACM on Interactive, Mobile, Wearable and Ubiquitous Technologies}} \bibinfo{volume}{8}, \bibinfo{number}{4} (\bibinfo{year}{2024}), \bibinfo{pages}{1--29}.
\newblock


\bibitem[Yang et~al\mbox{.}(2025)]%
        {yang2025contextagent}
\bibfield{author}{\bibinfo{person}{Bufang Yang}, \bibinfo{person}{Lilin Xu}, \bibinfo{person}{Liekang Zeng}, \bibinfo{person}{Kaiwei Liu}, \bibinfo{person}{Siyang Jiang}, \bibinfo{person}{Wenrui Lu}, \bibinfo{person}{Hongkai Chen}, \bibinfo{person}{Xiaofan Jiang}, \bibinfo{person}{Guoliang Xing}, {and} \bibinfo{person}{Zhenyu Yan}.} \bibinfo{year}{2025}\natexlab{}.
\newblock \showarticletitle{ContextAgent: Context-Aware Proactive LLM Agents with Open-World Sensory Perceptions}.
\newblock \bibinfo{journal}{\emph{arXiv preprint arXiv:2505.14668}} (\bibinfo{year}{2025}).
\newblock


\bibitem[Yang and Zheng(2022)]%
        {yang2022deepear}
\bibfield{author}{\bibinfo{person}{Qiang Yang} {and} \bibinfo{person}{Yuanqing Zheng}.} \bibinfo{year}{2022}\natexlab{}.
\newblock \showarticletitle{Deepear: Sound localization with binaural microphones}.
\newblock \bibinfo{journal}{\emph{IEEE Transactions on Mobile Computing}} \bibinfo{volume}{23}, \bibinfo{number}{1} (\bibinfo{year}{2022}), \bibinfo{pages}{359--375}.
\newblock


\bibitem[Yang et~al\mbox{.}(2024a)]%
        {yang2024maf}
\bibfield{author}{\bibinfo{person}{Yongjie Yang}, \bibinfo{person}{Tao Chen}, \bibinfo{person}{Yujing Huang}, \bibinfo{person}{Xiuzhen Guo}, {and} \bibinfo{person}{Longfei Shangguan}.} \bibinfo{year}{2024}\natexlab{a}.
\newblock \showarticletitle{MAF: Exploring mobile acoustic field for hand-to-face gesture interactions}. In \bibinfo{booktitle}{\emph{Proceedings of the 2024 CHI Conference on Human Factors in Computing Systems}}. \bibinfo{pages}{1--20}.
\newblock


\bibitem[Zhang et~al\mbox{.}(2024a)]%
        {zhang2023llamaadapter}
\bibfield{author}{\bibinfo{person}{Renrui Zhang}, \bibinfo{person}{Jiaming Han}, \bibinfo{person}{Chris Liu}, \bibinfo{person}{Peng Gao}, \bibinfo{person}{Aojun Zhou}, \bibinfo{person}{Xiangfei Hu}, \bibinfo{person}{Shilin Yan}, \bibinfo{person}{Pan Lu}, \bibinfo{person}{Hongsheng Li}, {and} \bibinfo{person}{Yu Qiao}.} \bibinfo{year}{2024}\natexlab{a}.
\newblock \bibinfo{title}{LLaMA-Adapter: Efficient Fine-tuning of Language Models with Zero-init Attention}.
\newblock
\showeprint[arxiv]{2303.16199}~[cs.CV]
\urldef\tempurl%
\url{https://arxiv.org/abs/2303.16199}
\showURL{%
\tempurl}


\bibitem[Zhang et~al\mbox{.}(2024b)]%
        {zhang2024earsavas}
\bibfield{author}{\bibinfo{person}{Xiyuxing Zhang}, \bibinfo{person}{Yuntao Wang}, \bibinfo{person}{Yuxuan Han}, \bibinfo{person}{Chen Liang}, \bibinfo{person}{Ishan Chatterjee}, \bibinfo{person}{Jiankai Tang}, \bibinfo{person}{Xin Yi}, \bibinfo{person}{Shwetak Patel}, {and} \bibinfo{person}{Yuanchun Shi}.} \bibinfo{year}{2024}\natexlab{b}.
\newblock \showarticletitle{The EarSAVAS Dataset: Enabling Subject-Aware Vocal Activity Sensing on Earables}.
\newblock \bibinfo{journal}{\emph{Proceedings of the ACM on Interactive, Mobile, Wearable and Ubiquitous Technologies}} \bibinfo{volume}{8}, \bibinfo{number}{2} (\bibinfo{year}{2024}), \bibinfo{pages}{1--26}.
\newblock


\bibitem[Zhu et~al\mbox{.}(2021)]%
        {zhu2021localizing}
\bibfield{author}{\bibinfo{person}{Hongzi Zhu}, \bibinfo{person}{Yuxiao Zhang}, \bibinfo{person}{Zifan Liu}, \bibinfo{person}{Xiao Wang}, \bibinfo{person}{Shan Chang}, {and} \bibinfo{person}{Yingying Chen}.} \bibinfo{year}{2021}\natexlab{}.
\newblock \showarticletitle{Localizing acoustic objects on a single phone}.
\newblock \bibinfo{journal}{\emph{IEEE/ACM Transactions on Networking}} \bibinfo{volume}{29}, \bibinfo{number}{5} (\bibinfo{year}{2021}), \bibinfo{pages}{2170--2183}.
\newblock


\bibitem[Zhu et~al\mbox{.}(2023)]%
        {zhu2023motionbert}
\bibfield{author}{\bibinfo{person}{Wentao Zhu}, \bibinfo{person}{Xiaoxuan Ma}, \bibinfo{person}{Zhaoyang Liu}, \bibinfo{person}{Libin Liu}, \bibinfo{person}{Wayne Wu}, {and} \bibinfo{person}{Yizhou Wang}.} \bibinfo{year}{2023}\natexlab{}.
\newblock \showarticletitle{Motionbert: A unified perspective on learning human motion representations}. In \bibinfo{booktitle}{\emph{Proceedings of the IEEE/CVF International Conference on Computer Vision}}. \bibinfo{pages}{15085--15099}.
\newblock


\end{thebibliography}


\end{document}